\documentclass[12pt]{iopart}
\usepackage{iopams}  
\usepackage{graphicx}  

\newcommand{\bfepsilon}{\boldsymbol{\epsilon}}
\newcommand{\sign}{\mathrm{sign}}
\newcommand{\calpha}{c_\alpha}
\newcommand{\salpha}{s_\alpha}
\newcommand{\cbeta}{c_\beta}
\newcommand{\sbeta}{s_\beta}
\newcommand{\cgamma}{c_\gamma}
\newcommand{\sgamma}{s_\gamma}
\newcommand{\bfr}{{\mathbf{r}}}
\newcommand{\bfk}{{\mathbf{k}}}
\newcommand{\bfn}{{\mathbf{n}}}
\newcommand{\bfe}{{\mathbf{e}}}
\newcommand{\bfa}{{\mathbf{a}}}
\newcommand{\bfb}{{\mathbf{b}}}
\newcommand{\bfc}{{\mathbf{c}}}
\newcommand{\bfd}{{\mathbf{d}}}
\newcommand{\bfu}{{\mathbf{u}}}
\newcommand{\bfv}{{\mathbf{v}}}
\newcommand{\id}{{\mathrm{Id}}}
\newcommand{\ee}{{\mathrm{e}}}

\newcommand{\sixj}[6]{\left\{ \begin{array}{@{}c@{\;}c@{\;}c@{}}
         #1 & #2 & #3 \\
         #4 & #5 & #6 
         \end{array}\right\}}
\newcommand{\potimes}{{\scriptstyle\otimes}}

\begin{document}

\title[Site symmetry and crystal symmetry]
{Site symmetry and crystal symmetry: a spherical tensor analysis}%

\author{Christian Brouder, Am\'elie Juhin, Am\'elie Bordage,
Marie-Anne Arrio}
%  Christos Gougoussis}

\address{Institut de Min\'eralogie et de Physique des Milieux
Condens\'es,
CNRS UMR 7590, Universit\'es Paris 6 et 7, IPGP, 140 rue de Lourmel,
75015 Paris, France.  }
\ead{christian.brouder@impmc.jussieu.fr}
\begin{abstract}
The relation between the properties of a specific crystallographic site
and the properties of the full crystal is discussed by using spherical
tensors. The concept of spherical tensors is introduced and the way
it transforms under the symmetry operations of the site and from
site to site is described in detail. The law of spherical tensor coupling
is given and illustrated with the example of the electric dipole and
quadrupole transitions in x-ray absorption spectroscopy. 
The main application of the formalism is the reduction of
computation time in the calculation of the properties of crystals
by band structure methods. The general approach is illustrated by
the examples of substitutional chromium in spinel
and substitutional vanadium in garnet.
\end{abstract}

Uncomment for PACS numbers title message
\maketitle

\section{Introduction}
This paper deals with the relation between the property of
specific sites in a crystal and the property of the crystal 
itself. We shall discuss in particular the case of
x-ray absorption spectroscopy, but many of our results are
general.

The aim of this paper is to introduce x-ray physicists,
experimentalists and theoreticians, to the use of spherical
tensors to analyze theoretical and experimental spectra.
Therefore, our presentation is as pedagogical as we can manage.
The manipulation of spherical tensors can be
quite complicated, we have tried to reduce it to
the simplest possible rules.

An atom in a crystal occupies generally a crystallographic
site that is not unique. The present paper was inspired by
the concrete case of a
chromium atom sustituted for aluminum in spinel
\cite{Juhin}. 
The chromium atom can occupy sixteen equivalent trigonal sites 
in the cubic cell.
The (normalized) x-ray absorption spectrum of chromium is 
the average of the spectra of the chromium atoms occupying
the sixteen sites. The angular-dependent spectrum of a chromium
atom in a specific site is determined by its environment,
but some features of this spectrum are lost by the average
over the equivalent sites. 
If we want to investigate the deformation of the environment due
to the substitution of chromium for aluminium, we need to know
the spectrum of chromium at a given site. But some features
of this spectrum are washed out by the fact that we can only measure
an average spectrum. The present paper describes precisely the
relation between the spectrum of a given site and the spectrum
of the crystal. 

If we want now to calculate the spectrum of chromium, we need
to put a chrominum atom at a specific aluminum site, 
relax the environment, carry out a self-consistent calculation 
with and without core hole, and calculate the spectrum of this site.
Then, we need to repeat the procedure for all equivalent sites
and take the average spectrum. A considerable amount of time
can be saved if we calculate the spectrum of one site and
deduce the spectrum of the other sites by symmetry considerations.
The present paper gives all the tools required to do so.

Let us first defend the case of spherical tensors, that will
be precisely defined in the next section.
Most physical properties are tensors and they are usually
taken to be Cartesian tensors. In this paper, we want to
stress the advantage of using spherical tensors. 
They are a refinement with respect to Cartesian tensor.
For example, a second-rank Cartesian tensor is the sum
a zeroth-rank, a first-rank and a second-rank  spherical tensors.
This refinement enables us to discard irrelevant contributions.
For example, the quadrupole transitions in x-ray absorption
spectroscopy are represented by a fourth-rank Cartesian tensor
with 81 components, whereas it is represented by a zeroth-rank,
a second-rank and a fourth-rank spherical tensor with 15 components
altogether. Moreover, the spherical average used to represent
powder sample is obtained by taking only one spherical tensor component
(the zeroth-rank tensor) whereas it is the sum of twelve
Cartesian components. Thus, using spherical tensors can
save computing time.

The drawback of the spherical tensor analysis is
that some subtelties enter its practical use (the same is true
for Cartesian tensors).
A purpose of this paper is to give a detailed
presentation of these problems and of their solutions.

In practice, one faces the frustrating task of sorting
out the various conventions used by the authors
and to determine whether the 
rotation is active or passive, if the row index of 
the Wigner matrix increases from left to right as in
a normal matrix or decrease as in reference \cite{BL}, p.~47,
which definition of the Wigner matrix is used
(seven inequivalent definitions are found in the literature
\cite{VMK}), how the tensor transforms under complex conjugation, etc.
Misprints are also numerous in the literature and we
made thorough efforts to give correct formulas.

The paper starts with a presentation of the Cartesian and spherical
tensors and a detailed description of the notation. Then,
the concept of tensor coupling is introduced and simple
formulas are given to change the coupling order. 
These methods are applied to the calculation of electric
dipole and quadrupole transitions.
This completes the generalities on spherical tensors.
Then, we consider the case of crystals.
We show how the symmetries of a crystal site
constrains the spherical tensor components
in the reference frame of the site and in the reference
frame of the crystal. We also give the precise formula 
relating these reference frames. As an illustration,
we treat the example of the Al site in spinel and garnet.
Finally, we determine the spherical tensor describing
the full crystal from those of the sites. 
An appendix gathers the formulas that were used to
calculate rotation matrices, Wigner matrices and 
solid harmonics.

\section{Cartesian and spherical tensors}
\label{cartsphsect}
A Cartesian tensor is the generalization of a scalar, a vector
or a matrix. A vector $\bfr=(x,y,z)$ is transformed by a rotation
$R$ into a vector $\bfr'=R\bfr$, so that
$\bfr'_i=\sum_{j=1}^3 R_{ij}\bfr_j$. A vector is also called
a first-rank tensor. A matrix $A$ defined by
$A_{ij}=\bfr_i\bfr_j$ transforms under rotation into
\begin{eqnarray}
A'_{ij} &=& \bfr'_i\bfr'_j = \sum_{lm} R_{il}R_{jm} A_{lm}.
\label{rang2}
\end{eqnarray}
Any matrix $A_{ij}$ that transforms under rotation as
in equation (\ref{rang2}) is called a second-rank Cartesian tensor.
More generally, an $n$th-rank Cartesian tensor is a generalized matrix
$A_{i_1\dots i_n}$ that transforms under rotation as in equation
(\ref{rang2}), but with the product of $n$ matrices $R$ instead
of just two. Cartesian tensors are ubiquitous in physics. 
As we shall see, the electric dipole transitions are described by
a second-rank tensor, the quadrupole ones by a fourth-rank
tensor.

Cartesian tensors have very simple transformation rules under
rotation, but they suffer from a severe drawback: they are
not irreducible. To see what that means, consider a 
second-rank tensor $A_{ij}$. Its trace is 
$t=\sum_{i=1}^3 A_{ii}$. It transforms under rotation into
\begin{eqnarray*}
t' &=& \sum_{i=1}^3 A'_{ii} =
\sum_{ilm} R_{il}R_{im} A_{lm} =
\sum_{lm} \delta_{lm} A_{lm} =t,
\end{eqnarray*}
where we used the fact that $RR^T=\id$, where
$R^T$ is the transpose of $R$. We recover the fact that the
trace of a matrix is invariant: it transforms into itself by
rotation. 
A second-order Cartesian tensor gives another interesting object,
the vector $\bfv$, which is defined by
$\bfv_i=\sum_{jk} \epsilon_{ijk} A_{jk}$, where
the sum is over $j$ and $k$ from 1 to 3. The Levi-Civita symbol
$\epsilon_{ijk}$ is 1 if $(i,j,k)$ is a cyclic permutation
of $(1,2,3)$, it is -1 if $(i,j,k)$ is another permutation of
$(1,2,3)$ and it is 0 if two indices are identical.
Then, using the identity
$\sum_{jk} \epsilon_{ijk} R_{jl} R_{km} = \sum_j R_{ij}\epsilon_{jlm}$,
we see that $\bfv$ transforms under rotation as a vector:
$\bfv'=R\bfv$. 
Therefore, from a second-rank Cartesian tensor, we can
build a linear combination of its elements (the trace) that is invariant
under rotation, and three linear combinations of its elements
that transform into each other as the components of a vector. 
More generally, a tensor is said to be \emph{reducible}
when there are linear combinations of its elements that transform
into each other under rotation. When a tensor is not reducible,
it is called irreducible. Thus, a vector is irreducible but
a second-rank Cartesian tensor is reducible. 
The irreducible tensors are called
\emph{spherical tensors} and will be the main topic of
this paper.

The first spherical tensors were the spherical harmonics $Y^m_\ell$.
For each $\ell$, there are $2\ell+1$ spherical harmonics
$Y^m_\ell$ that transform into each other by rotation.
More precisely, for each rotation $R$, there is a unitary
matrix $D^\ell$, called a \emph{Wigner matrix} (to be precisely
defined in the next section), such that the
rotation of  $Y^m_\ell$ by $R$ is
\begin{eqnarray*}
RY_\ell^m &=& \sum_{m'=-\ell}^\ell Y_\ell^{m'} D^\ell_{m'm}(R).
\end{eqnarray*}
Spherical tensors are defined in analogy with spherical harmonics.
An $\ell$th rank spherical tensor, denoted by $T^\ell$, is a set
of $2\ell+1$ components, written $T^\ell_m$
where $m=-\ell,-\ell+1,\dots,\ell-1,\ell$, that transform under
rotation as
\begin{eqnarray}
RT^\ell_m &=& \sum_{m'=-\ell}^\ell T^\ell_{m'} D^\ell_{m'm}(R).
\label{RTL}
\end{eqnarray}
This definition is not very concrete, but we shall see how spherical
tensors are built in practice. It is an unfortunate but historical
fact that the position of $\ell$ and $m$ is different in
the spherical harmonics $Y_\ell^m$ and the spherical tensors
$T^\ell_m$.

When many spherical tensors are involved in a formula,
we use also the notation $P^a$, $Q^b$, etc.
For notational convenience, we shall often write
$RT^\ell =  T^\ell D^\ell(R)$ for equation (\ref{RTL}),
as for the product of a matrix and a vector.
Moreover, the product of two Wigner matrices 
will be written $D^\ell(R) D^\ell(R')$.

\subsection{Further symmetries}
Spherical tensors can satisfy other symmetries.
For example, most of them are built from Hermitian
operators and satisfy time-reversal symmetry:
(see equation~(4) p.~61 of \cite{VMK}).
\begin{eqnarray}
(T^\ell_m)^* &=& (-1)^m T^\ell_{-m}.
\label{TRS}
\end{eqnarray}
In this paper, the only tensors that do not satisfy
time-reversal symmetry are those built from the
polarization vector $\bfepsilon$ that can possibly
be complex. In that case we have
$\big(T^\ell_m(\bfepsilon)\big)^* = (-1)^m T^\ell_{-m}(\bfepsilon^*)$.
This happens for instance when $T^\ell$ is a solid harmonics
(see section \ref{solharmsect}). 

Inversion is another common symmetry. In pure electric dipole and quadrupole
transitions inversion does not play any role because all tensors
are even (i.e. invariant under inversion). However, the interference 
between dipole and quadrupole
transitions is always an odd spherical tensor 
\cite{GoulonM,BrouderNCD,Carra00}.

\section{Notation}
The many conventions and misprints found in the literature
forces us to describe in detail
our notation and conventions.

We consider \emph{active} rotation, that is rotation
that move the points and not the reference frame. For example
the rotation through an angle $\psi$ about the $z$-axis
is represented by
\begin{eqnarray*}
R_z(\psi) = \left( \begin{array}{ccc}
               \cos\psi & -\sin\psi & 0 \\
               \sin\psi & \cos\psi & 0 \\
               0 & 0 & 1
   \end{array}\right).
\end{eqnarray*}
After an active rotation $R$, the coordinates
$(\bfr_1,\bfr_2,\bfr_3)$ of the vector $\bfr$
are transformed into the coordinates
$\bfr'_i=\sum_j R_{ij}\bfr_j$ of
$\bfr'=R\bfr$.
In a \emph{passive} rotation, the reference frame is rotated:
the basis vectors $\bfe_i$ are transformed into the
basis vectors $\bfe'_i=\sum_j R_{ij}\bfe_j$.
Thus, the coordinates of a point $\bfr$ are
transformed by the inverse matrix:
$\bfr'=R^{-1}\bfr$.

To describe the transformation of the properties of a crystal
under rotation, we consider the case of its
charge density $\rho(\bfr)$. 
After a rotation changing $\bfr$ into
$\bfr'=R\bfr$, the charge density $\rho$ is transformed into
a ``rotated'' charge density $\rho'$ of the rotated crystal.
To determine $\rho'$, we require that the value of
the charge density is not modified by the rotation. More precisely,
we want that $\rho'(\bfr')=\rho(\bfr)$.
Therefore, the rotated function $\rho'$ is defined
by $\rho'(\bfr')=\rho(R^{-1}\bfr')$.
For later convenience, we denote the rotated function
$\rho'$ by $R\rho$. The use of the same symbol $R$ to denote
the rotation of both the vectors and the functions should not
bring too much confusion.
The presence of the inverse rotation $R^{-1}$ 
in the definition of $R\rho$ ensures that $R'(R\rho)=(R'R)\rho$
(see Ref.~\cite{LudwigFalter} p.~59).

\subsection{Wigner rotation matrices}
We denote by $D^{\ell}_{m'm}(R)$ the Wigner rotation
matrix corresponding to the rotation $R$.
For example,
$D^\ell_{m'm}\big(R_z(\psi)\big)=\delta_{mm'} \ee^{-im\psi}$.
The Wigner rotation matrices define a unitary representation
of the rotation group, so that
\begin{eqnarray}
D^{\ell}_{m'm}(R^{-1}) &=& \big(D^{\ell}_{mm'}(R)\big)^*,
\label{unitaire}
\end{eqnarray}
and 
\begin{eqnarray}
D^{\ell}_{m'm}(RR') &=& \sum_{m''=-\ell}^\ell
    D^{\ell}_{m'm''}(R)D^{\ell}_{m''m}(R').
\label{grouprep}
\end{eqnarray}

\subsection{Spherical harmonics}
The spherical harmonics are defined by
(see Ref.~\cite{BL} p.~68)
\begin{eqnarray}
Y_\ell^m(\theta,\phi) &=& \sqrt{\frac{2\ell+1}{4\pi}} 
   \big(D^\ell_{m0}(R_{\theta\phi})\big)^*,
\label{defspharm}
\end{eqnarray}
where $R_{\theta\phi}$ is the rotation defined by the
Euler angles $(\phi,\theta,0)$. 
For notational convenience, we denote by 
$\bfn$ the vector
$(\sin\theta\cos\phi,\sin\theta\sin\phi,\cos\theta)$
and we write $Y_\ell^m(\bfn)$ and $R_\bfn$ for
$Y_\ell^m(\theta,\phi)$ and $R_{\theta\phi}$.
This notation is justified by the fact that
$Y_\ell^m(\bfn)$ can be defined for any vector 
$\bfr$, not necessarily normalized. The resulting
functions are called \emph{solid harmonics} and are 
described in section \ref{solharmsect}. 
Solid harmonics are required, for example, in the presence of
elliptically polarized x-rays because $\bfn$ has
then complex coordinates.
A three-dimensional Cartesian vector 
$\bfr=(x,y,z)$ can be turned into 
a set of three solid harmonics
\begin{eqnarray}
Y_1^{-1}(\bfr) &=& \sqrt{\frac{3}{8\pi}}(x-iy),\\
Y_1^{0}(\bfr) &=& \sqrt{\frac{3}{4\pi}}z,\\
Y_1^{1}(\bfr) &=& -\sqrt{\frac{3}{8\pi}}(x+iy).
\label{solidharm}
\end{eqnarray}

This definition of spherical harmonics implies
\begin{eqnarray*}
Y_\ell^m(R\bfn) &=& \sum_{m'} Y_\ell^{m'}(\bfn) D^\ell_{m'm}(R^{-1}).
\end{eqnarray*}
This is proved by noticing that the argument
$R\bfn$ of the spherical harmonics corresponds to the argument
$RR_\bfn$ of the Wigner matrix in equation (\ref{defspharm}). 
By equation (\ref{unitaire}), we have
$\big(D^\ell_{m0}(RR_{\bfn})\big)^*=
D^\ell_{0m}\big((RR_\bfn)^{-1}\big)$.
The result follows from $(RR_\bfn)^{-1}=R_\bfn^{-1}R^{-1}$
and the group representation property, equation (\ref{grouprep}).
The same property is true for solid harmonics.
Therefore,
\begin{eqnarray}
(RY_\ell^m)(\bfr) &=& Y_\ell^m(R^{-1}\bfr)
= \sum_{m'} Y_\ell^{m'}(\bfr) D^\ell_{m'm}(R).
\label{RYellm}
\end{eqnarray}
The presence of the spherical harmonics on the left of
the Wigner rotation matrices ensures that $R'(RY_\ell^m) = (R'R)Y_\ell^m$.
To show this, multiply equation (\ref{RYellm}) on the left by
$R'$:
\begin{eqnarray*}
\big(R'(RY_\ell)\big)(\bfr) &=& 
(R'Y_\ell)(\bfr) D^\ell(R) =Y_\ell(\bfr) D^\ell(R')D^\ell(R)
\\&=&
Y_\ell(\bfr) D^\ell(R'R) =(R'R)Y_\ell(\bfr).
\end{eqnarray*}
In the foregoing proof, we simplified the notation by omitting
the component index $m$ (as described at the end of section \ref{cartsphsect}).

\section{Building tensor operators}
Physical properties can be represented by spherical tensors
that can often be built by coupling lower rank tensors.
We illustrate this construction by the example of electric
dipole and quadrupole transitions.
We shall use the remarkable toolbox for spherical tensor calculations
elaborated by Varshalovich, Moskalev and Khersonskii \cite{VMK}.

Many spherical tensors used in physics are obtained
by coupling vectors. 
A three-dimensional Cartesian vector 
$\bfv=(x,y,z)$ can be turned into a first-rank spherical
tensor $\bfv^1$ by defining 
\begin{eqnarray}
\bfv^1_{-1} &=& (x-iy)/\sqrt{2},\nonumber\\
\bfv^1_{0}  &=& z,\nonumber\\
\bfv^1_{1}  &=& -(x+iy)/\sqrt{2}.
\label{defbfv1}
\end{eqnarray}
Note that solid harmonics $Y_\ell^m(\bfv)$ 
are also spherical tensors built from $\bfv$
and that $\bfv^1=Y_1(\bfv)\sqrt{4\pi/3}$.
However, the factor $\sqrt{4\pi/3}$ is cumbersome
and the definition $\bfv^\ell=Y_\ell(\bfv)\sqrt{4\pi/(2\ell+1)}$
is often preferred.

Two spherical tensors $P^a$ and $Q^b$ of ranks $a$
and $b$ can be \emph{coupled} into a spherical tensor
of rank $c$, denoted by $\{P^a\potimes Q^b\}^c$, and defined by
\begin{eqnarray*}
\{P^a\potimes Q^b\}^c_\gamma &=&
\sum_{\alpha=-a}^a \sum_{\beta=-b}^b
(a\alpha b\beta|c\gamma) P^a_\alpha Q^b_\beta,
\end{eqnarray*}
where $(a\alpha b\beta|c\gamma)$ are Clebsch-Gordan coefficients
\cite{BL,VMK}. 
The Clebsch-Gordan coefficient is zero if $\gamma\not=\alpha+\beta$.
The triangle relation implies $|a-b|\le c \le a+b$.
The notation $\{P^a\potimes Q^b\}^c$ is inspired by
ref. \cite{VMK}, except for the fact that they write
the rank as an index instead of an exponent.
For example, the coupling of two vectors (i.e. first-rank
spherical tensors) gives a zeroth-rank,
a first-rank and a second-rank spherical tensors.
The zeroth-rank tensor obtained by coupling two
vectors is proportional to the scalar product of
these vectors:
$\{\bfu^1\potimes\bfv^1\}^0=-\bfu\cdot\bfv/\sqrt{3}$, because
$(1\alpha 1-\alpha|00)=-(-1)^\alpha/\sqrt{3}$. More generally,
we define the \emph{scalar product} of two spherical
tensors $P^a$ and $Q^a$ of the same rank to be
(see \cite{VMK}, p.~64 and 65)
\begin{eqnarray}
P^a \cdot Q^a &=& \sum_{\alpha=-a}^a (-1)^\alpha
  P^a_{-\alpha} Q^a_{\alpha} = (-1)^a \sqrt{2a+1}
  \{P^a\potimes Q^a\}^0.
\label{scalaire}
\end{eqnarray}

It is often necessary to modify the coupling order of the
tensors. For example, when calculating electric dipole
transitions, we have to calculate $|\langle f | \bfepsilon\cdot\bfr|i\rangle|^2$,
where $\bfr$ is coupled to $\bfepsilon$ by the
scalar product and then multiplied by its complex conjugate.
As we shall see in the next section, it is more convenient from 
the physical point of view to couple the x-ray polarization vectors 
$\bfepsilon$ and $\bfepsilon^*$.
For that purpose, there is a powerful identity
\begin{eqnarray}
\{P^a\potimes Q^a\}^0 \cdot \{R^d\potimes S^d\}^0
&=& 
 \sum_g 
  (-1)^g
\frac{\{P^a\potimes R^d\}^g \cdot \{Q^a\potimes S^d\}^g}
{\sqrt{(2a+1)(2d+1)}},
\label{recouple0}
\end{eqnarray}
where $g$ runs from $|a-d|$ to $a+d$ by the triangle relation.  
This identity is proved in section
\ref{couplingidsect}.

In the next two sections, we illustrate the recoupling
methods with the calculation of electric dipole and
quadrupole transitions.
Similar methods were used to investigate the
interference of electric and quadrupole transitions
and their interferences \cite{BrouderNCD,Carra00,Carra03}
or to calculate x-ray scattering cross-sections
\cite{MarriCarra}.

\subsection{Dipole}
The electric dipole transitions amplitudes are 
given by the formula
$T_{fi}=\langle f | \bfepsilon\cdot\bfr |i \rangle$.
If we denote 
$\langle f |\bfr |i \rangle$ by $\bfr_{fi}$, 
equation (\ref{scalaire}) gives us
$T_{fi}=\bfepsilon\cdot \bfr_{fi}=-\sqrt{3} 
\{\bfepsilon^1\potimes\bfr^1_{fi}\}^0$.
For notational convenience, we remove the exponent 1
in the spherical tensors $\bfepsilon^1$ and $\bfr^1_{fi}$.
This should not bring confusion: if a vector takes part in a coupling, 
it is a first-rank tensor.
Using the recoupling identity (\ref{recouple0}), we find
the dipole transition intensity
\begin{eqnarray}
|T_{fi}|^2 &=& 3  \{\bfepsilon^*\potimes\bfr^*_{fi}\}^0
    \{\bfepsilon\potimes\bfr_{fi}\}^0
=
 \sum_{a=0}^2 (-1)^a \{\bfepsilon^*\potimes\bfepsilon\}^a\cdot
    \{\bfr_{fi}^*\potimes\bfr_{fi}\}^a.
\label{dipole}
\end{eqnarray}
Note that, for elliptic polarization, $\bfepsilon$ is complex,
whereas the wavevector $\bfk$ is always real.
Each term of a decomposition over spherical tensors
has often a clear physical meaning. 
In equation (\ref{dipole}), the variables concerning the
incident x-ray (i.e. $\bfepsilon$ and $\bfepsilon^*$)
are gathered in 
 $\{\bfepsilon^*\potimes\bfepsilon\}^a$,
the variables concerning the crystal are in
   $\{\bfr_{fi}^*\potimes\bfr_{fi}\}^a$.
Thus, we can easily investigate the influence of a
rotation $R$ of the crystal on the absorption
cross-section
\begin{eqnarray*}
R|T_{fi}|^2 &=&
 \sum_{a=0}^2 (-1)^a \{\bfepsilon^*\potimes\bfepsilon\}^a\cdot
    \big(R\{\bfr_{fi}^*\potimes\bfr_{fi}\}^a\big)
\\&=&
 \sum_{a=0}^2 (-1)^a \{\bfepsilon^*\potimes\bfepsilon\}^a\cdot
    \big(\{\bfr_{fi}^*\potimes\bfr_{fi}\}^a D^a(R)\big),
\end{eqnarray*}
where we used equation (\ref{RTL}) and that 
$\{\bfr_{fi}^*\potimes\bfr_{fi}\}^a$ is a $a$th-rank spherical tensor.
In particular, the spectrum of a powder sample is given
by the average over all orientations, i.e. over all
rotations $R$.
This average is very simple with spherical tensors
$\langle D^a(R) \rangle = \delta_{a,0}$.
Thus, the term $a=0$ gives the spectrum of a powder,
called the isotropic spectrum.
\begin{eqnarray*}
\langle|T_{fi}|^2\rangle &=&
  \{\bfepsilon^*\potimes\bfepsilon\}^0\cdot
    \{\bfr_{fi}^*\potimes\bfr_{fi}\}^0
= \frac{1}{3} (\bfepsilon^*\cdot\bfepsilon)\,
  (\bfr_{fi}^*\cdot\bfr_{fi})
= \frac{|\bfr_{fi}|^2}{3},
\end{eqnarray*}
where we used equation (\ref{scalaire}) and 
$|\bfepsilon|^2=1$.

To interpret the term $a=1$, we use the relation
between vectors $\bfu$, $\bfv$ and the
corresponding first-rank tensors
$\bfu^1$, $\bfv^1$ (we restore the tensor rank
in $\bfu^1$, $\bfv^1$ for clarity):
according to equation~(\ref{vecprod}),
$\{\bfu^1\potimes\bfv^1\}^1$ is the first-rank
tensor corresponding to the vector
$(i/\sqrt{2}) \bfu\times\bfv$.
Therefore,
\begin{eqnarray*}
 \{\bfepsilon^*\potimes\bfepsilon\}^1\cdot
    \{\bfr_{fi}^*\potimes\bfr_{fi}\}^1
&=& -\frac{1}{2}
  (\bfepsilon^*\times \bfepsilon)\cdot 
  (\bfr_{fi}^*\times\bfr_{fi}).
\end{eqnarray*}
The first cross product is related to 
the rate of circular polarization 
$P_c$ and to wavevector direction
$\hat{k}$ of the incident x-ray by
$\bfepsilon^*\times \bfepsilon=-i P_c \hat{k}$
\cite{BrouderNCD}.
Moreover, the second cross product is
zero for a non-magnetic sample because
time-reversal symmetry implies 
$\bfr_{fi}^*=\bfr_{fi}$.
Therefore, the term $a=1$ describes
magnetic circular dichroism.

The term $a=2$ describes 
the linear dichroism of x-ray spectra.
The number of non-zero components of 
$\{\bfr_{fi}^*\potimes\bfr_{fi}\}^2$
depends on the symmetry of the crystal
\cite{BrouderAXAS}. We shall determine the number
of non-zero components in the case of spinel and garnet.

\subsection{Quadrupole}
We consider the case of electric quadrupole transitions,
which is more involved than dipole transitions.
We start from the quadrupole transition operator 
$T=\bfepsilon\cdot\bfr \, \bfk\cdot\bfr$ and we rewrite
it in terms of spherical tensors using equation (\ref{prod0}):
$T = 3 \{\{\bfepsilon\potimes\bfr\}^0\potimes\{\bfk\potimes\bfr\}^0\}^0$.
In this expression $\bfepsilon$ is coupled with $\bfr$
and $\bfk$ with $\bfr$. As in the electric dipole, 
we want to gather all the terms concerning the crystal into
a single tensor. For that purpose, we use
equation (\ref{recouple0}) with the sum over $g$
changed into a sum over $a$
\begin{eqnarray*}
T &=& 
\sum_{a=0}^2 (-1)^a
\{\bfepsilon\potimes\bfk\}^a\cdot\{\bfr\potimes\bfr\}^a.
\end{eqnarray*}
The term $a=0$ is zero because, according to equation 
(\ref{scalaire}),
$\{\bfepsilon\potimes\bfk\}^0=-(1/\sqrt{3}) \bfepsilon\cdot\bfk =0$ 
since the polarization and wavevectors are perpendicular.
The term $a=1$ is zero because equation (\ref{vecprod}) 
gives us
$\{\bfr\potimes\bfr\}^1=(i/\sqrt{2}) \bfr\times\bfr =0$.
Thus, $T$ is reduced to the single term
\begin{eqnarray*}
T &=& \{\bfepsilon\potimes\bfk\}^2\cdot\{\bfr\potimes\bfr\}^2
= \sqrt{5} 
  \{\{\bfepsilon\potimes\bfk\}^2\potimes\{\bfr\potimes\bfr\}^2\}^0.
\end{eqnarray*}
The tensor $\{\bfr\potimes\bfr\}^2$ can be
expressed in terms of spherical harmonics
(equation~(23), p.~67 of \cite{VMK})
\begin{eqnarray*}
\{\bfr\potimes\bfr\}^2_m &=&
\sqrt{\frac{8\pi}{15}} Y_2^m(\bfr) =
\sqrt{\frac{8\pi}{15}} r^2 Y_2^m(\theta,\phi),
\end{eqnarray*}
where $r$, $\theta$ and $\phi$ are the spherical coordinates
of $\bfr$.
For completeness, we give the components of $\{\bfepsilon\potimes\bfk\}^2$:
\begin{eqnarray*}
\{\bfepsilon\potimes\bfk\}^2_{\pm 2} &=& 
   \frac{(\epsilon_x\pm i\epsilon_y)(k_x\pm ik_y)}{2},\\
\{\bfepsilon\potimes\bfk\}^2_{\pm 1} &=& \mp
   \frac{(\epsilon_x\pm i\epsilon_y)k_z+\epsilon_z(k_x\pm ik_y)}{2},\\
\{\bfepsilon\potimes\bfk\}^2_{0} &=& 
  =\frac{3\epsilon_z k_z - \bfepsilon\cdot\bfk}{\sqrt{6}}=
  \sqrt{\frac{3}{2}} \epsilon_zk_z.
\end{eqnarray*}

The electric quadrupole transition intensities are proportional
to $|T_{fi}|^2$, where the transition amplitude is
$T_{fi}=\langle f | T | i\rangle$. Therefore,
$|T_{fi}|^2 = 5\{P^2\potimes Q^2\}^0 \{R^2\potimes S^2\}^0$,
with
$P^2=\{\bfepsilon^*\potimes\bfk\}^2$,
$Q^2=\langle f|\{\bfr\potimes\bfr\}^2|i\rangle^* $,
$R^2=\{\bfepsilon\potimes\bfk\}^2$,
$S^2=\langle f|\{\bfr\potimes\bfr\}^2|i\rangle$,
\begin{eqnarray}
|T_{fi}|^2 &=& 
5\{P^2\potimes Q^2\}^0 \{R^2\potimes S^2\}^0
=\sum_{a=0}^4 (-1)^a
\{P^2\potimes R^2\}^a \cdot \{Q^2\potimes S^2\}^a.
\label{quadrupole}
\end{eqnarray}
The apparent simplicity of this calculation is essentially
due to the powerful tools given in reference \cite{VMK}. 
A straightforward approach is quite heavy \cite{BrouderAXAS}.
If the sample is nonmagnetic or the x-ray polarization is
linear, the terms $a=1$ and $a=3$ are zero.

As in the electric dipole case, the term $a=0$ corresponds
to the isotropic spectrum obtained by measuring a powder.
Equation (\ref{prod4}) gives us 
$\{P^2\potimes R^2\}^0=k^2/(2\sqrt{5})$ and the isotropic spectrum is
\begin{eqnarray*}
\langle |T_{fi}|^2 \rangle &=&
 k^2\frac{\{Q^2\potimes S^2\}^0}{2\sqrt{5}}.
\end{eqnarray*}
The calculation of this average will be discussed
in section \ref{sphavsect}.

If $P^2=R^2$ (i.e. the x-rays are linearly polarized)
or $Q^2=S^2$ (i.e. the sample is non magnetic), then
the terms $a=1$ and $a=3$ are zero.
More generally, for any tensor $T^a$ with integer rank $a$,
$\{T^a\potimes T^a\}^c$ is zero for $c$ odd.
This is due to the symmetry of the Clebsch-Gordan 
coefficients~\cite{BL,VMK}
$(b\beta a\alpha|c\gamma)=(-1)^{a+b-c}
(a\alpha b\beta|c\gamma)$:
\begin{eqnarray*}
\{T^a\potimes T^a\}^c_\gamma &=& \sum_{\alpha,\beta}
(a\alpha a \beta|c \gamma)  T^a_{\alpha} T^a_{\beta}
= (-1)^{2a-c} \sum_{\beta,\alpha}
(a\alpha a \beta|c \gamma) T^a_{\beta} T^a_{\alpha}
\\&=&  (-1)^{c} \{T^a\potimes T^a\}^c_\gamma,
\end{eqnarray*}
where we first exchanged the summation variables
$\alpha$ and $\beta$,
and used the symmetry of the Clebsch-Gordan coefficients,
then the commutativity of $T^a_{\alpha}$ and $T^a_{\beta}$,
and the fact that $a$ is an integer.

If we consider only the case of linearly polarized x-rays
or non-magnetic samples, only the terms $a=0$,
$a=2$ and $a=4$ are possibly non zero. 
The number of independent components depends on the
crystal symmetry and is tabulated in \cite{BrouderAXAS}.

\section{Site symmetry}
We describe how to calculate the spherical tensor
of a crystallographic site, assuming that it is
invariant under the symmetries of the site.
We can work in a reference frame corresponding either to
the site symmetries or to the crystal symmetries.
If we take the aluminum sites of spinels as an example,
the site symmetry is $D_{3d}$ whereas the crystal is cubic.
In the reference frame adapted to the site symmetry,
the $z$ axis is along a diagonal of the cube. 
In the reference frame adapated to the crystal, 
the $z$ axis is along an edge of the cube. 
The symmetrized tensors has usually less nonzero components
in the site axes, but they are easier to calculate in the
crystal axes. We also give the way to go from one reference
frame to the other.
Two examples are treated in detail: a spinel and a garnet.

\subsection{Symmetrized tensor}
If $G$ is the symmetry group of the crystal,
then a site has a symmetry group $G'$ which is a
subgroup of $G$. To know the general form of 
a spherical tensor invariant under the 
site symmetries, we start from a general tensor
$T^\ell$ and we calculate 
the symmetrized tensor $\langle T^\ell_m \rangle$ compatible
with the site symmetries by using the classical formula
\begin{eqnarray}
\langle T^\ell_m \rangle &=& \frac{1}{|G'|} 
  \sum_{R'} \sum_{m'=-\ell}^\ell T^\ell_{m'} D^\ell_{m'm}(R'),
\label{siteaverage}
\end{eqnarray}
where $R'$ runs over the symmetry operations of the subgroup
$G'$ and where $|G'|$ is the number of elements of $G'$.
From the physical point of view, this formula means that the
tensor $\langle T^\ell_m \rangle$ is obtained by
averaging over all the symmetry operations that leave the
site invariant. From the mathematical point of view,
we project onto the subset of tensors that are invariant
by any symmetry operation of $G'$. To check this,
take any operation $R$ in $G'$ and evaluate the action
of $R$ on the symmetrized tensor:
\begin{eqnarray*}
\langle RT^\ell_m \rangle &=& 
   \frac{1}{|G'|} \sum_{R'} \sum_{m'm''} 
      T^\ell_{m''}D^\ell_{m''m'}(R)D^\ell_{m'm}(R')
\\&=&
   \frac{1}{|G'|} \sum_{R'} \sum_{m''} 
      T^\ell_{m''}D^\ell_{m''m}(RR')=\langle T^\ell_m \rangle,
\end{eqnarray*}
because, $G'$ being a group, the set of operations $RR'$
where $R'$ runs over $G'$ is the same as the set of operations
of $G'$.

\subsection{Site and crystal axes}
The rotations $R$ can be expressed either in the site axes
or in the crystal axes. We shall see in the examples that
the symmetric tensors are simpler in the site axes.
Moreover, some computer programs need to be used in the
site axes.\footnote{The most prominent example is the series
of multiplet programs written by Cowan, Butler, Thole, Ogasawara
and Searle\cite{Cowan,Butler,Thole85,Kotani,Kuiper}.}
However, the matrices $R$ are easier to determine in
the crystal axes, because they correspond to the
symmetry operations of the crystal that leave the site
invariant.
Both cases will be treated in the examples of the following
sections.

It is also necessary to describe precisely how to go from
one reference frame to the other.
If $\bfe_1$, $\bfe_2$, $\bfe_3$ are the orthonormal
axes associated with the crystal and 
$\bfe'_1$, $\bfe'_2$, $\bfe'_3$ those of the site,
we denote by $R$ the rotation matrix such that
$\bfe'_i=\sum_j R_{ij} \bfe_j$.
For example, if $\bfe_i$ are the axes of the cube
and $\bfe'_i$ are the trigonal axes along the
$(-1,1,1)$ direction, then 
$R$ is the rotation matrix corresponding to the Euler angles
$(0,\arccos(1/\sqrt{3}),\pi/4)$. It is the inverse of the rotation 
matrix of equation (\ref{Butler43}). 
It can be checked that $\bfe'_3=(-\bfe_1+\bfe_2+\bfe_3)/3$
(i.e. the three-fold axis is the $z$-axis of the site along the 
$(-1,1,1)$ direction of the cube) and $\bfe'_2=(\bfe_1+\bfe_2)/\sqrt{2}$
(i.e.  the $y$ axis of the site is along the $(1,1,0)$ direction
of the cube).
If $R'$ is a symmetry operation in the cubic axes of the cube leaving
the site invariant, the basis change formula in a vector space implies
that $RR'R^{-1}$ is a symmetry operation in the 
trigonal axes of the site.
A spherical tensor $T^\ell$ will be denoted by 
$T^\ell(3)$ when it is expressed in the trigonal axes and by
$T^\ell(4)$ when it is expressed in the cubic axes. The argument
3 and 4 mean that the $z$ axis is along a three-fold axis for a trigonal
basis and a four-fold axis for a cubic basis. The relation between
$T^\ell(3)$ and $T^\ell(4)$ is given by the formula
\begin{eqnarray}
T^\ell_m(4) &=& \sum_{m'} T^\ell_{m'}(3) D^\ell_{m'm}(R).
\label{passe34}
\end{eqnarray}

We give now two examples.

\subsection{The example of spinel}
We illustrate this method with the example of the aluminium site in
spinel MgAl$_2$O$_4$, which is the most common mineral of the spinel
structural family.  The spinel structure is derived from a
face-centred-cubic close-packing of oxygens with a  space group
symmetry $Fd\bar{3}m$. 
The conventional cubic cell contains 8 formula units, i.e. 32
oxygen atoms with 24 cations in tetrahedral and 
pseudo-octahedral interstices.
With origin choice 2 \cite{Hahn}, the Mg$^{2+}$ cations occupy 8 
tetrahedral sites, which are located at the
special 8a Wyckoff positions (1/8,1/8,1/8), with
$\bar{4}3m$ ($T_d$) point symmetry.
The Al$^{3+}$ cations occupy 16 pseudo-octahedral sites at
the special 16d Wyckoff positions (0,1/4,3/4), with
$\bar{3}m$ ($D_{3d}$) point symmetry. This symmetry corresponds to a small
elongation of the octahedron along the trigonal axis, arising from a
small departure of  the position of the oxygens from the perfect fcc
arrangement. The O$^{2-}$ ions are located at the Wyckoff positions
32e (u,u,u) with point symmetry $3m$.

\subsubsection{The site frame}
The simplest results are obtained when the reference
frame of the site is used. The point group of the
site we consider is $D_{3d}$.
The group $D_{3d}$ has six pure rotations and 
the same six rotations multiplied by the inversion.
We assume that the property that
we investigate is not sensitive to the inversion operator,
so that we only have to consider the six pure rotations.
It is natural to take the $z$-axis along the three-fold axis and
the $y$ axis along one of the $C_2$ axes. The results do not
depend on which $C_2$ axis is chosen. However, they would
be different if the $y$ axis were chosen, for example,
between two $C_2$ axes. 
The six pure rotations are: the unit,
the $C_3$ rotation  about the $z$-axis through the angle $2\pi/3$,
its square $C_3^2$, the $C_2$ rotation about the $y$-axis through
the angle $\pi$ and the other two rotations
$C_3C_2$ and $C_3^2C_2$.
These rotations have Euler angles $(0,0,0)$, $(0,0,2\pi/3)$,
$(0,0,4\pi/3)$, $(0,\pi,0)$, $(0,\pi,4\pi/3)$ and 
$(0,\pi,2\pi/3)$, respectively.
These rotations will be denoted by $R_1,\dots,R_6$.

To calculate the symmetrized tensors for this
site, we use equation (\ref{siteaverage}). 
The special cases given in section \ref{EulerWignersect}
enable us to show that
$D^\ell_{m'm}(R_1)=\delta_{m'm}$,
$D^\ell_{m'm}(R_2)=\delta_{m'm}\ee^{-2m i \pi/3}$ and
$D^\ell_{m'm}(R_3)=\delta_{m'm}\ee^{-4m i\pi/3}$.
Therefore, the sum $\sum_{j=1}^3 D^\ell_{m'm}(R_j)$
is $3\delta_{m'm}$ if $m$ is an integer multiple of 3, and zero
otherwise. We calculate the Wigner matrices for the other
three rotations and we obtain
\begin{eqnarray*}
\frac{1}{|G'|} \sum_{R'} D^\ell_{m'm}(R') &=&
\frac{1}{6} \sum_{j=1}^6 D^\ell_{m'm}(R_j) =
  \frac{\delta_{m'm} + (-1)^{\ell-m} \delta_{m,-m'}}{2},
\end{eqnarray*}
if $m$ and $m'$ are integer multiples of 3, and zero otherwise.
Equation (\ref{siteaverage}) is then applied to a general
fourth-rank tensor $T^4_m(3)$, where the argument $(3)$ denotes
the trigonal axes and we obtain the non-zero components
of the symmetrized tensor $\langle T^4_m(3) \rangle$
\begin{eqnarray*}
\langle T^4_0(3) \rangle &=& T^4_0(3),\\
\langle T^4_3(3) \rangle &=& -\langle T^4_{-3}(3) \rangle
= \frac{T^4_3(3)- T^4_{-3}(3)}{2}.
\end{eqnarray*}
For the second-rank tensor, all symmetrized components
are zero, except for $\langle T^2_0(3) \rangle = T^2_0(3)$.
Of course, we have also the relation
$\langle T^0_0(3) \rangle = T^0_0(3)$, which is valid
for any group.
Now we show that time-reversal symmetry implies that the
symmetrized tensors are real.
According to equation (\ref{TRS})
$(T^\ell_0)^*=T^\ell_0$, so that $T^\ell_0$ is real.
Still by equation (\ref{TRS}) we have
$(T^\ell_{3})^*=-T^\ell_{-3}$. Thus,
$\langle T^4_3(3) \rangle^*=(-T^4_{-3}(3)+ T^4_{3}(3))/2
=\langle T^4_3(3) \rangle$ is real as well.

\begin{figure}[ht]
\begin{center}
\includegraphics[width=10cm]{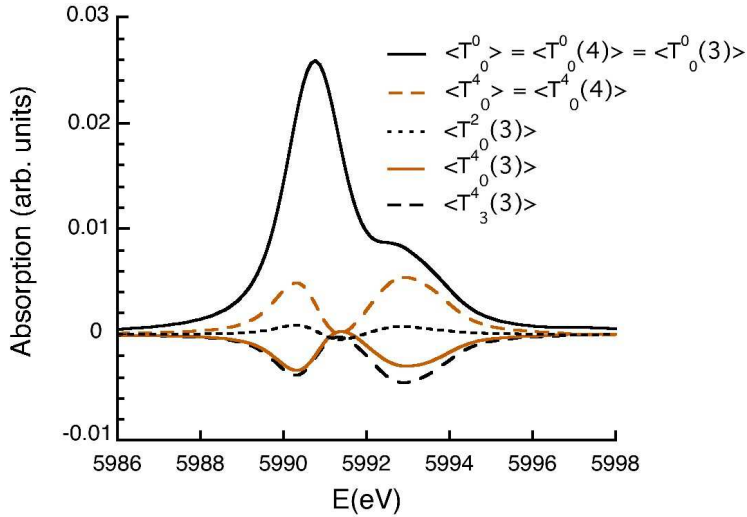}
\caption{\label{fig:spinel}The symmetrized tensors
$\langle T^0_0(3)\rangle$, $\langle T^2_0(3)\rangle$,
$\langle T^4_0(3)\rangle$ and $\langle T^4_3(3)\rangle$
in the site frame, for the electric quadrupole transitions a
the K-edge of a chromium atom substituted for
aluminium in spinel. After averaging over the sites,
only the symmetrized tensors 
$\langle T_0^0\rangle$ and $\langle T_0^4\rangle$ remain
as independent parameters.}
\end{center}
\end{figure}

In x-ray absorption spectra, the symmetrized tensors
are spectral functions depending on the photon energy.
For the example of the electric quadrupole transitions
we take, for each energy $\hbar\omega$,
\begin{eqnarray*}
T^\ell &=& \pi^2 \alpha_0 \sum_f 
\{\langle f|\{\bfr\potimes\bfr\}^2|i\rangle^*\potimes
\langle f|\{\bfr\potimes\bfr\}^2|i\rangle\}^\ell
\delta(E_f-E_i-\hbar\omega),
\end{eqnarray*}
where $\alpha_0$ is the fine structure constant,
$E_i$ and $E_f$ the energy of the initial and
final states.
The symmetrized tensors $\langle T^\ell(3)\rangle$
can be calculated by multiplet programs.
The value of these tensors for a chromium atom substituted
for aluminum in spinel is given in figure~\ref{fig:spinel}
(see reference \cite{Juhin} for the details of the calculation)

\subsubsection{The crystal frame}
\label{crystframesect}
We consider now the same average in the crystal axes.
The chosen Al site with reduced coordinates 
(0,1/4,3/4) has a three-fold axis along the (-1,1,1)
direction and a two-fold axis along the (0,1,0) direction.
Therefore, the six pure rotations of $D_{3d}$
are now (i) the identity, denoted by $(x,y,z)$,
(ii) a $C_3$ rotation  about (-1,1,1),  
denoted by $(-y,z,-x)$, (iii)
its square $(-z,-x,y)$, (iv) a rotation of $\pi$
about (1,1,0) denoted by $(y,x,-z)$, (v) a rotation of $\pi$
about (1,0,1) denoted by $(z,-y,x)$, (vi) a rotation of $\pi$
about (0,1,-1) denoted by $(-x,-z,-y)$.
The notation used for the rotations is the result of
the operation $R\bfr$ in the cubic axes. For example, the $C_3$
rotation gives 
\begin{eqnarray*}
R\bfr =  \left( \begin{array}{ccc}
               0 & -1 &  0 \\
               0 &  0 & 1 \\
              -1 &  0 &  0
   \end{array}\right)
\left( \begin{array}{c}
               x \\
               y \\
               z 
   \end{array}\right)
=
\left( \begin{array}{c}
              -y \\
               z \\
              -x 
   \end{array}\right).
\end{eqnarray*}
The corresponding Euler angles are
$(0,0,0)$, $(\pi/2,\pi/2,0)$, $(\pi,\pi/2,\pi/2)$,
$(0,\pi,\pi/2)$, $(0,\pi/2,\pi)$ and $(3\pi/2,\pi/2,3\pi/2)$.

We apply again equation (\ref{siteaverage}) to the general
second-rank tensor $T^2_m(4)$, where the argument
$(4)$ stands for the cubic axes. This gives us
the symmetrized tensor $\langle T^2(4)\rangle$
\begin{eqnarray}
\langle T^2_0(4)\rangle &=& 0,\nonumber\\
\langle T^2_{-2}(4)\rangle &=& \langle T^2_{2}(4) \rangle^* = -i\lambda,
\nonumber\\
\langle T^2_{-1}(4)\rangle &=&-\langle T^2_{1}(4) \rangle^* = (1+i) \lambda,
\label{spinelT2}
\end{eqnarray}
with
\begin{eqnarray*}
\lambda &=& \frac{
\Im{T^2_2(4)}- \Re{T^2_1(4)}+\Im{T^2_{1}(4)}}{3},
\end{eqnarray*}
where we have used time-reversal symmetry as in
equation (\ref{TRS}).
Note that $\lambda$ is real.
For the tensor $\langle T^4(4)\rangle$,
\begin{eqnarray}
\langle T^4_0(4)\rangle &=&
\sqrt{\frac{14}{5}} \langle T^4_4(4)\rangle =
\sqrt{\frac{14}{5}}\langle T^4_{-4}(4)\rangle^*=\xi,\nonumber\\
\langle T^4_{-3}(4)\rangle &=& -\langle T^4_{3}(4) \rangle^* 
=(1-i)\sqrt{7} \zeta ,
\nonumber\\
\langle T^4_{-2}(4)\rangle &=&\langle T^4_{2}(4)\rangle^* = 2i\sqrt{2} \zeta,
\nonumber\\
\langle T^4_{-1}(4)\rangle &=& -\langle T^4_{1}(4) \rangle^* = (1+i) \zeta,
\label{spinelT4}
\end{eqnarray}
with 
\begin{eqnarray*}
\xi &=& \frac{7 T^4_0(4) + \sqrt{70} \Re T^4_4(4)}{12},\\
\zeta &=& -\frac{
\sqrt{7}(\Re{T^4_3(4)}+\Im{T^4_{3}(4)})+2\sqrt{2}\Im{T^4_2(4)}
   +\Re{T^4_1(4)}-\Im{T^4_{1}(4)}}{24}.
\end{eqnarray*}
Note that $\xi$ and $\zeta$ are real.

\subsubsection{From site to crystal frame}

From this example, it is clear that the symmetrized tensor
$\langle T^\ell(3) \rangle$ in the site frame is much simpler
the same tensor $\langle T^\ell(4) \rangle$ in the crystal frame.
The relation between the trigonal and cubic axes in
Butler's tables is worked out in section \ref{Butlersect}
of the Appendix.

To go from one to the other we apply equation (\ref{passe34}) and
we obtain the relations
\begin{eqnarray}
\langle T^4_{0}(4) \rangle &=& -\frac{7\langle  T^4_{0}(3)\rangle
   + 2\sqrt{70} \langle T^4_{3}(3) \rangle}{18},
   \label{T4043}\\
\langle T^4_{-2}(4) \rangle &=& i\frac{\sqrt{10}\langle T^4_{0}(3)\rangle
   - \sqrt{7} \langle T^4_{3}(3) \rangle}{9}.
   \label{T4243}
\end{eqnarray}
We recover the relations (\ref{spinelT4}) with 
\begin{eqnarray*}
\zeta &=& \frac{2\sqrt{5}\langle  T^4_{0}(3)\rangle
   - \sqrt{14} \langle T^4_{3}(3) \rangle}{36},\\
\xi &=& -\frac{7\langle  T^4_{0}(3)\rangle
   + 2\sqrt{70} \langle T^4_{3}(3) \rangle}{18}.
\end{eqnarray*} 

For the second-rank tensor we find the relations (\ref{spinelT2})
with $\lambda=-\langle  T^2_{0}(3)\rangle/\sqrt{6}$.
For the zeroth-rank tensor we have obviously
$\langle  T^0_{0}(4)\rangle=\langle  T^0_{0}(3)\rangle$.

\subsection{The example of garnet}
We consider now the Al site in garnet with the example of grossular
Ca$_3$Al$_2$(SiO$_3$)$_4$, which is a cubic mineral, 
with the space group $Ia\bar{3}d$. 
The cubic cell contains 96 oxygen, 24 calcium, 24 silicium and 16
aluminium atoms. The Al$^{3+}$ cations are at the 16a Wyckoff 
positions (0,0,0). This site is a
slightly distorted octahedral, with a small elongation along the
(111) axis of the cube, and has the $\hat{3}$ ($C_{3i}$) point symmetry.

We calculate the symmetrized tensor in the 
site frame as for spinel, but with the smaller
symmetry group $C_{3i}$.

\begin{figure}[ht]
\begin{center}
\includegraphics[angle=0,width=10cm]{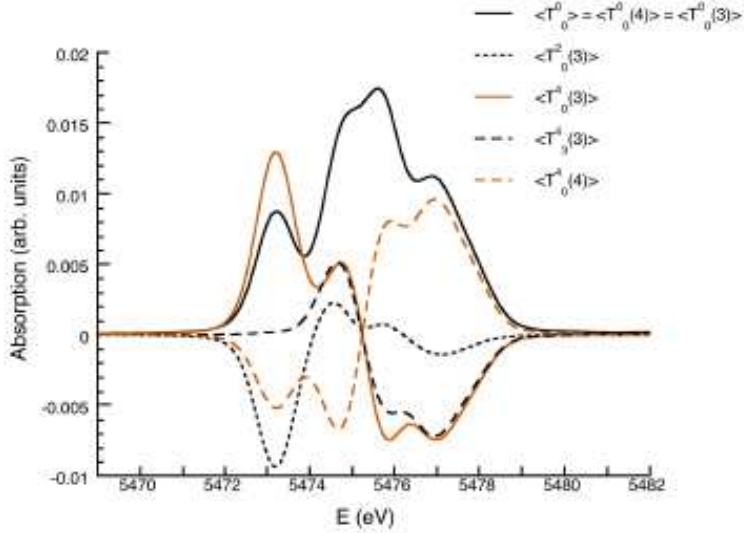}
\caption{\label{fig:garnet}The symmetrized tensors
$\langle T^0_0(3)\rangle$, $\langle T^2_0(3)\rangle$,
$\langle T^4_0(3)\rangle$ and $\langle T^4_3(3)\rangle$
in the site frame, for the electric quadrupole transitions at the
K-edge of a vanadium atom substituted for
aluminium in grossular garnet.}
\end{center}
\end{figure}

We find that the non-zero tensor components are~\cite{BrouderAXAS}
$\langle T^4_{-3}(3) \rangle$, $\langle T^4_{0}(3) \rangle$,
$\langle T^4_{3}(3) \rangle$, $\langle T^2_{0}(3) \rangle$
and $\langle T^0_{0}(3) \rangle$,
as illustrated in figure \ref{fig:garnet}
(see reference \cite{Bordage} for the details of the calculation).

It would not be very illuminating to calculate directly the
symmetrized tensor in the crystal frame. It is more
interesting to rotate the tensor.
Indeed, in the group $C_{3i}$, the $z$-axis is specified by
the rotation axis, but the $y$-axis is arbitrary in the
plane perpendicular to the rotation axis.
This arbitrariness can be quite useful. For instance,
the parametrization of the Hamiltonian
is simplified by choosing the $y$ axis so that
a crystal field parameter is set to zero (see Ref.~\cite{Butler} p.~184).
This simplifies the calculation of the eigenstates but the
parameter reappears as the angle $\alpha$ between the $y$ axis 
and the $(1,1,0)$ direction of the cube in the plane
perpendicular to the $(-1,1,1)$ direction.

The corresponding rotation matrix is
\begin{eqnarray*}
R &=& \sqrt{\frac{2}{3}} \left( \begin{array}{ccc}
               \cos(\alpha+\pi/3) &
               \cos(\alpha+2\pi/3) &
               \cos\alpha \\
               \sin(\alpha+\pi/3) &
               \sin(\alpha+2\pi/3) &
               \sin\alpha \\
               -1/\sqrt{2} &
               1/\sqrt{2} &
               1/\sqrt{2}
   \end{array}\right).
\end{eqnarray*}
The Euler angles are $\alpha$, $\arccos(1/\sqrt{3})$, $\pi/4$.
The angle $\alpha$ describes a rotation about the axis
(-1,1,1). Therefore, the $\alpha$ dependency of the result
is very simple: we have
$T^\ell_m(\alpha)=T^\ell_m(0)\ee^{-mi\alpha}$
because the corresponding Wigner matrix is
$D^\ell_{m'm}=\delta_{mm'} \ee^{-im\alpha}$.

To calculate the symmetrized tensor
$\langle T^4_m(4)\rangle$ in the cubic axes,
we use equation (\ref{passe34}), we put $s=\langle T^4_0(3) \rangle$ and
$t_r+it_i=\ee^{-i\alpha} \langle T^4_3(3) \rangle$
and we obtain
\begin{eqnarray}
\langle T^4_{-4}(4)\rangle &=& \langle T^4_{4}(4) \rangle^* =
   -\frac{\sqrt{70} s +20 t_r -12 i \sqrt{3} t_i}{36},\nonumber\\
\langle T^4_{-3}(4)\rangle &=& -\langle T^4_{3}(4) \rangle^* =
   (1-i)\frac{2\sqrt{35} s -7\sqrt{2} t_r +3 i \sqrt{6} t_i}{36},
\nonumber\\
\langle T^4_{-2}(4)\rangle &=& \langle T^4_{2}(4) \rangle^* = 
   i\frac{\sqrt{10} s -\sqrt{7} t_r}{9},
\nonumber\\
\langle T^4_{-1}(4)\rangle &=&-\langle T^4_{1}(4) \rangle^* =
   (1+i)\frac{2\sqrt{5} s -\sqrt{14} t_r +3 i \sqrt{42} t_i}{36},
\nonumber\\
\langle T^4_{0}(4)\rangle &=& 
   -\frac{7 s +2\sqrt{70} t_r}{18}.
\label{grenatT4}
\end{eqnarray}

The symmetrized second-rank tensor in the cubic frame
is the same as for spinel.

\begin{figure}[ht]
\begin{center}
\includegraphics[angle=0,width=10cm]{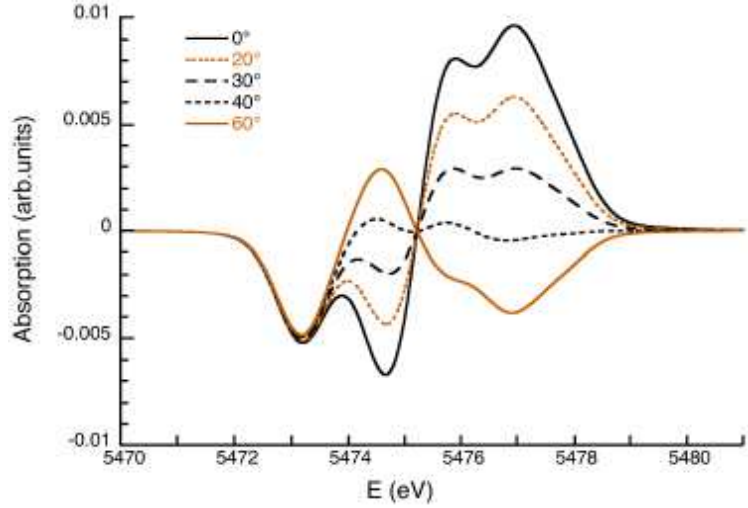}
\caption{\label{fig:angles}The crystal symmetrized tensor
$\langle T^4_0\rangle$ as a function of $\alpha$,
for the electric quadrupole transitions at the K-edge of
a vanadium atom substituted for
aluminium in grossular garnet.}
\end{center}
\end{figure}

The effect of the angle $\alpha$ on the experimental
spectrum can be considerable, as is illustrated
in figure \ref{fig:angles}.

\section{From site symmetry to crystal symmetry}
We consider in this section another type of problem.
We assume that we have calculated a symmetrized
tensor $\langle T^\ell\rangle$ for a certain site $A$.
We want to know the value of the same tensor
for all the sites equivalent to $A$.
In the first section, we describe how this can
be done. 
In an x-ray absorption measurement, we measure
the sum of the signals coming from all sites of the
crystal. Therefore, we do not really need to know
the spectrum of each site, we need to know the 
spectrum averaged over all the sites.
Two methods will be presented to do so: the coset method
and the brute force method. Finally, we treat the
examples of spinel and garnet.

\subsection{Changing site}
\label{changingsitesect}
In general, the symmetries of the crystal make
several equivalent sites. More precisely, if a 
crystal has the symmetry group $G$ and the site has
symmetry group $G'$, then the number of equivalent sites
in the crystal is the ratio $|G|/|G'|$.
Assume that we have calculated
a physical property described by a spherical tensor 
$T^\ell_m$ for a given site $A$. We want to calculate the
properties of the equivalent site $B$.

If $\bfr_A$ and $\bfr_B$ are the position vectors of site $A$
and $B$, there is a symmetry operation $R$ of $G$ such that
$\bfr_B=R \bfr_A$. If we denote by $G_A$ and $G_B$
the symmetry group of sites $A$ and $B$, then
the symmetry operation $R'$ of $G_A$ will be transformed
into the operation $RR'R^{-1}$ of $G_B$.
This can be checked because $R$ transforms the
rotation axis $\bfn$ of an operation $R'$ of $G_A$
into the axis $R\bfn$, and $RR'R^{-1}$
describes the rotation through the same angle as $R'$
but about the axis $R\bfn$. This is indeed a symmetry operation
of site $B$.
Moreover, the tensor $T^\ell_m(B)$ at site $B$ is
related to the tensor $T^\ell_m(A)$ at site $A$
by the relation 
\begin{eqnarray*}
T^\ell_m(B) &=& \sum_{m'=-\ell}^\ell T^\ell_{m'}(A)
    D^\ell_{m'm}(R^{-1}).
\end{eqnarray*}
We are now facing a typical subtelty of crystal symmetries.
We could have expected the argument of the Wigner matrix to be
$R$ instead of $R^{-1}$, but this is not the case because
the rotation $R$ is in fact a passive operation from the 
point of view of site $B$. The rotation transports the rotation
axis, which is a vector of the reference frame. 
By moving the atoms of the crystal, the rotation transports
the reference frame of the site. Therefore, we are in the
passive point of view and we need to use $R^{-1}$ because
our convention uses the active point of view.

To check this, calculate the symmetrized tensor around $B$.
\begin{eqnarray*}
\langle T^\ell_m(B) \rangle &=& \frac{1}{|G_B|} 
  \sum_{R'_B\in G_B} \sum_{m'=-\ell}^\ell T^\ell_{m'}(B)
D^\ell_{m'm}(R'_B).
\end{eqnarray*}
Now, we can use the fact that the operations $R'_B$ of $G_B$
can be obtained from the operations $R'_A$ of $G_A$
by $R'_B=RR'_AR^{-1}$. Therefore
\begin{eqnarray*}
\langle T^\ell_m(B) \rangle &=& \frac{1}{|G_A|} 
  \sum_{R'_A\in G_A} \sum_{m'=-\ell}^\ell T^\ell_{m'}(B) 
  D^\ell_{m'm}(RR'_AR^{-1}),
\end{eqnarray*}
where we used $|G_A|=|G_B|$, a consequence of the isomorphism 
between $G_A$ and $G_B$.
We see that this is only compatible with the transformations
$T^\ell(B)=T^\ell(A)D^\ell(R^{-1})$ and
$\langle T^\ell(B)\rangle =\langle T^\ell(A)\rangle D^\ell(R^{-1})$.

\subsection{Changing the x-ray beam}
If we have calculated the spectrum of site $A$,
it is possible to obtain the spectrum of any site equivalent to $A$
by calculating the spectrum of site $A$ for a rotated x-ray beam.
This is physically clear because, if you rotate both the crystal
and the x-ray, the spectrum does not change. 
Thus, if site $A$ is measured with a polarization $\bfepsilon$
and a wavevector $\bfk$, the spectrum of the site obtained by 
rotating $A$ with rotation matrix $R$ is the
same as the spectrum of $A$ measured with a polarization $R^{-1}\bfepsilon$
and a wavevector $R^{-1}\bfk$. To be fully convinced, 
we prove that, if $P^a$ and $Q^a$ are two $a$th-rank tensors,
then  $P^a\cdot (RQ^a)=(R^{-1} P^a)\cdot Q^a$. Equations
(\ref{RTL}) and (\ref{scalaire}) give us
\begin{eqnarray*}
P^a\cdot (RQ^a) &=& P^a\cdot (Q^aD^a(R))
=\sum_{\alpha\beta} (-1)^\alpha P^a_{-\alpha} Q^a_\beta D^a_{\beta\alpha}(R).
\end{eqnarray*}
The symmetry relation
(see ref.~\cite{VMK}, p.~79)
$D^a_{\beta\alpha}(R)=(-1)^{\beta-\alpha} D^a_{-\alpha-\beta}(R^{-1})$
enables us to write
\begin{eqnarray*}
P^a\cdot (RQ^a) &=& 
\sum_{\alpha\beta} (-1)^{\beta} P^a_{-\alpha} 
  D^a_{-\alpha-\beta}(R^{-1})Q^a_\beta
\\&=&
\sum_{\beta} (-1)^{\beta} \big(P^a D^a(R^{-1})\big)_{-\beta}Q^a_\beta
=(R^{-1} P^a)\cdot Q^a.
\end{eqnarray*}

For the example of the electric dipole transition
probability (\ref{dipole}) we find
\begin{eqnarray}
\{\bfepsilon^*\potimes\bfepsilon\}^a\cdot
    R\big(\{\bfr_{fi}^*\potimes\bfr_{fi}\}^a\big)
&=&
R^{-1}\big(\{\bfepsilon^*\potimes\bfepsilon\}^a\big)\cdot
    \{\bfr_{fi}^*\potimes\bfr_{fi}\}^a.
\label{Rdipole}
\end{eqnarray}
It remains to prove that the rotation of 
$\{\bfepsilon^*\potimes\bfepsilon\}^a$ corresponds to
the rotation of $\bfepsilon^*$ and $\bfepsilon$.
This is done by using the following identity:
\begin{eqnarray*}
R\{P^a\potimes Q^b\}^c &=& \{(RP^a)\potimes (RP^b)\}^c.
\end{eqnarray*}
To demonstrate the latter identity, we write the rotation
in terms of the Wigner matrices and
the coupled tensor in terms of the Clebsch-Gordan coefficients:
\begin{eqnarray}
R\{P^a\potimes Q^b\}^c_\gamma &=& 
\sum_{\alpha\beta\gamma'} 
(a\alpha b\beta|c\gamma') P^a_\alpha Q^b_\beta D^c_{\gamma'\gamma}(R).
\label{RPQ}
\end{eqnarray}
The classical identity (\ref{WignerCG}) transforms this expression
into
\begin{eqnarray*}
R\{P^a\potimes Q^b\}^c_\gamma &=& 
\sum_{\alpha\beta\alpha'\beta'} 
(a\alpha' b\beta'|c\gamma) P^a_\alpha Q^b_\beta 
D^a_{\alpha\alpha'}(R)
D^b_{\beta\beta'}(R)
\\&=&
\sum_{\alpha'\beta'} 
(a\alpha' b\beta'|c\gamma) (RP)^a_{\alpha'} (RQ)^b_{\beta'} 
=\{(RP^a)\potimes (RP^b)\}^c_\gamma.
\end{eqnarray*}

Equations (\ref{Rdipole}) and (\ref{RPQ}) yield
\begin{eqnarray*}
\{\bfepsilon^*\potimes\bfepsilon\}^a\cdot
    R\big(\{\bfr_{fi}^*\potimes\bfr_{fi}\}^a\big)
&=&
\{(R^{-1}\bfepsilon^*)\potimes(R^{-1}\bfepsilon)\}^a\cdot
    \{\bfr_{fi}^*\potimes\bfr_{fi}\}^a.
\end{eqnarray*}
In other words, rotating the crystal by $R$ gives the same
result as rotating the x-ray beam by $R^{-1}$.
The same result is true for the electric quadrupole transition
probabilities, except for the fact that the polarization
$\bfk$ are rotated by $R^{-1}$ and not only $\bfepsilon$.

\subsection{The coset method}
The coset method is a powerful way to calculate the 
the tensor symmetrized over the crystal from the tensor
symmetrized over a single site.
We first introduce some mathematical concepts.

\subsubsection{Mathematical aspects}
The symmetry group $G_A$ of a site $A$ is a subgroup of the
symmetry group $G$ of the crystal. The number of sites equivalent
to $A$ in the unit cell is the ratio $|G|/|G_A|$, where
$|G|$ and $|G_A|$ denote the number of elements of
$G$ and $G_A$. This ratio, denoted by $n$ in the following,
is an integer by the Euler-Lagrange theorem (\cite{LudwigFalter} p.~11).
We can pick up $n$ symmetry operations $g_i$
in $G$ that transport site $A$ to the $n$ sites
equivalent to $A$. 

We first introduce some mathematical concepts \cite{LudwigFalter}.
If $g$ is an element of a group $G$ and $G_A$ a subgroup of $G$, 
the set $gG_A=\{gh: h\in G_A\}$ is called a \emph{coset}.
If we take two elements $g$ and $g'$ of $G$, then
either $gG_A$ and $g'G_A$ are either identical or disjoint
(i.e. they have no element in common). 
The number of different cosets is $n=|G|/|G_A|$
and the number of elements in each coset is $|G_A|$.
Moreover, every element of $G$ belongs to one and only one coset. 
Therefore, if we pick up any element $g_i$
in each coset, we have $G= g_1 G_A \cup \dots \cup g_n G_A$ and
each $g_i$ is called a \emph{representative} of its
coset.

\subsubsection{Cosets in a crystal}
We apply these concepts to a crystal. The
crystal symmetry group is a space group $G$.
The symmetry group of a site $A$ is the set of operations
of $G$ that leave site $A$ invariant.
More precisely, if $\bfr_A$ is the coordinate vector of site $A$,
then $G_A=\{g\in G | g(\bfr_A)=\bfr_A\}$.
It is clear that $G_A$ is a subgroup of $G$.
If $n=|G|/|G_A|$, there are $n$ sites in the crystal
that are equivalent to $A$. The group $G$ is partitioned into
$n$ cosets $g_1G_A,\dots,g_nG_A$. All the elements of a
given coset send site $A$ to the \emph{same} equivalent site.
Take an element $g_i$ in each coset $g_i G_A$.

The symmetrized tensor $\langle T^\ell\rangle_X$ over the full crystal 
is obtained from the site-symmetrized tensor 
$\langle T^\ell\rangle_A$ by the operation
\begin{eqnarray}
\langle T^\ell\rangle_X &=& \frac{1}{n} \sum_{i=1}^n 
  \langle T^\ell\rangle_A D^\ell(g_i^{-1}),
\label{TellX}
\end{eqnarray}
where, for a space group operation $g$, 
$D^\ell(g)$ is the Wigner matrix corresponding to the proper 
rotation of $g$.
Equation (\ref{TellX}) can be described by saying that we average 
over the symmetries of site $A$, then we go to another sites with 
$g_i$ and we average over the symmetries of that other site.

We show now that equation (\ref{TellX}) gives the same result
as an average over all crystal symmety operations.
We already know that
\begin{eqnarray*}
\langle T^\ell\rangle_A &=& \frac{1}{|G_A|}\sum_{h\in G_A} 
  T^\ell D^\ell(h)
= \frac{1}{|G_A|}\sum_{h\in G_A} 
  T^\ell D^\ell(h^{-1}),
\end{eqnarray*}
because the sum over the elements
of a group is the same as the sum over the inverse elements of this group.
Therefore
\begin{eqnarray*}
\langle T^\ell\rangle_X &=& \frac{1}{n|G_A|} \sum_{i=1}^n \sum_{h\in G_A}
  T^\ell D^\ell(h^{-1})D^\ell(g_i^{-1})
=  \frac{1}{|G|}
 \sum_{i=1}^n \sum_{h\in G_A}
  T^\ell D^\ell(h^{-1}g_i^{-1})
\\&=&
  \frac{1}{|G|} \sum_{i=1}^n \sum_{h\in G_A}
  T^\ell D^\ell((g_ih)^{-1})
= \frac{1}{|G|}  \sum_{g\in G}
  T^\ell D^\ell(g),
\end{eqnarray*}
where we used $(g_ih)^{-1}=h^{-1}g_i^{-1}$ and $|G|=n|G_A|$.
Note that the proof holds because we used sets
$g_i G_A$ (i.e. left cosets) and not
$G_A g_i$ (i.e. right cosets).

In other words, the average over
the site symmetries followed by the average over the
sites gives the average over the crystal symmetries.
This can be considered as a factorization of the average, 
because $\sum_{g} D^\ell(g)=\sum_h D^\ell(h)\sum_i D^\ell(g_i^{-1})$.
It can be checked easily that the result of equation
(\ref{TellX}) is the same if we replace $g_i$ by any
$g'_i$ in $g_iG_A$.
We illustrate this by our two favorite examples,
spinel and garnet.

\subsubsection{The example of spinel}
\label{exspinelsite}
We call site $A$ the site (0,1/4,3/4) of the spinel structure.
The space group $G$ has 192 operations\footnote{The full space group
has of course an infinite number of operations because it contains
all the translations by a vector of the Bravais lattice. We consider
here the group $G$ obtained as the quotient of the full space
group by the group of translations of the simple cubic lattice. 
We could have used the quotient of the full space group by the group
of translations of the fcc lattice (which has only 48 operations),
however our choice is simpler from the programming point of view:
we take the symmetry operations given by the table \cite{Hahn},
and we identify two points whose coordinates differ by an integer.}.
Twelve of them leave 
site A invariant:
$(x,y,z)$, 
$(-y+3/4,z,-x+3/4)$,
$(-z+3/4,-x+3/4,y)$, 
$(y+1/4,x+3/4,-z+1/2)$,
$(z+1/4,-y+1/2,x+3/4)$,
$(-x,-z+1/2,-y+1/2)$,
$(-x,-y+1/2,-z+1/2)$,
$(y+1/4,-z+1/2,x+3/4)$, 
$(z+1/4,x+3/4,-y+1/2)$, 
$(-y+3/4,-x+3/4,z)$, 
$(-z+3/4,y,-x+3/4)$, 
and $(x,z,y)$.
This set of twelve operations is a group
isomorphic to $D_{3d}$.
The isomorphism $\varphi$ is described explicitly 
as follows. If $\bfr_A$ is the coordinate vector of
site $A$, for any operation $g$ of the
set, we define the operation $\varphi(g)$ by
$\varphi(g)(\bfr)=g(\bfr+\bfr_A)-\bfr_A$.
It is an isomorphism because
$\varphi(g'g)=\varphi(g')\varphi(g)$:
\begin{eqnarray*}
\varphi(g')\varphi(g)(\bfr) &=& \varphi(g')\big(g(\bfr+\bfr_A)-\bfr_A\big)
= (g'g)(\bfr+\bfr_A)-\bfr_A=\varphi(g'g)(\bfr),
\end{eqnarray*}
and it can be checked that the image by $\varphi$ of the
first six operations are the rotations listed in section 
\ref{crystframesect}, the image of the other six operations
are the same rotations multiplied by -1.
The space group operations $g$ are selected by the condition
that the site is fixed: $g(\bfr_A)=\bfr_A$.
Thus, the origin is a fixed point of $\varphi(g)$:
$\varphi(g)(0)=g(\bfr_A)-\bfr_A=0$.

There are 16 sites equivalent to $A$ because
the ratio $|Fd\bar{3}m|/|D_{3d}|$ is 16. However, each site is equivalent
to three other sites by pure lattice translations
$(1/2,1/2,0)$, $(1/2,0,1/2)$ and $(0,1/2,1/2)$. The x-ray spectrum
of these sites will be equal because their orientations with respect
to the x-ray beam are the same. Therefore, we are left with
4 equivalent sites: $A$ itself and the sites with coordinates 
$(1/4,3/4,0)$, $(3/4,0,1/4)$ and $(1/2,1/2,1/2)$. 
A representative of the coset corresponding to each of these sites is
$(-y+1/2,x+3/4,z+1/4)$, $(-x+3/4,-y+3/4,z)$ and $(y+1/4,-x,z+1/4)$.

\subsubsection{The example of garnet}
We consider now site $A$ as the  $Al$ site in grossular 
garnet with the three-fold axis along the $(-1,1,1)$ direction.
Its coordinates are (1/2,1/2,0). It is left
invariant by the following six operations:
identity: $(x,y,z)$, rotation through $2\pi/3$ about the
$(-1,1,1)$ axis: $(-y,z+1/2,-x+1/2)$, rotation through $4\pi/3$ about the
$(-1,1,1)$ axis: $(-z+1/2,-x,y+1/2)$, and the same operations
multiplied by an inversion $(-x,-y,-z)$,
$(y,-z+1/2,x+1/2)$ and $(z+1/2,x,-y+1/2)$.
This group is $C_{3i}$. The number of cosets (i.e. the number
of sites equivalent to $A$) is $|Ia\bar{3}d|/|C_{3i}|=16$. 
If we remove the lattice translation (1/2,1/2,1/2) we are
left with 8 equivalent sites.

The approach in terms of cosets is quite powerful in practice
because it completely avoids the explicit description of the
sites and of the symmetry operations that transform a site
into another one. Many programs compute the symmetry operations
of the sites (for example PWSCF). It is then enough to take one of
these sites, to choose any representative $g_i$ in each coset
and to calculate the contribution of all equivalent sites by 
the formula $\langle T^\ell\rangle_A D^\ell(g_i^{-1})$.
The average over the crystal is then obtained with equation
(\ref{TellX}). 

\subsection{The brute force method}
If one is not interested in the contribution of each
site to the spectrum of the crystal, a  still simpler 
solution is to take the average of $\langle T^\ell\rangle_A$
over all the symmetry operations of the crystal. 
This is not very clever because the site operations
have already been taken into account and we average over
them a second time, but this method is well suited
for computers.

We show now that averaging the site-symmetrized tensor
over all the symmetry operations of the crystal gives
the same result as the coset method.
\begin{eqnarray*}
\frac{1}{|G|}  \sum_{g\in G}
\langle T^\ell\rangle_A D^\ell(g) &=&
\frac{1}{|G_A||G|}   \sum_{h\in G_A} \sum_{g\in G}
T^\ell D^\ell(h) D^\ell(g) 
\\&=&
\frac{1}{n|G_A|^2}   \sum_{h,h'\in G_A} \sum_{i}
T^\ell D^\ell(h) D^\ell(h') D^\ell(g_i^{-1}) 
\\&=&
\frac{1}{n|G_A|}   \sum_{h\in G_A} \sum_{i}
T^\ell D^\ell(h) D^\ell(g_i^{-1}) 
=\langle T^\ell \rangle_X,
\end{eqnarray*}
where we used the identity
$\sum_{h,h'\in G_A}  D^\ell(hh')=|G_A|\sum_{h\in G_A}  D^\ell(h)$,
which is readily established.
Therefore, the average over all the symmetry operations of the
crystal gives the same result as the average over the sites,
irrespective of the number of equivalent sites.

\subsection{The case of spinel}
We illustrate the coset method with the case of spinel.
We first notice that the absorption cross section
is invariant under a translation of the Bravais lattice,
because such a translation multiplies the wavefunction
by a phase (independent of $\bfr$) that disappears
in the square modulus.
Therefore, by removing the translations, we can replace the 
representatives of the four cosets given in section \ref{exspinelsite} 
by the four rotations about the $z$-axis of the crystal
through angles 0, $\pi/2$, $\pi$ and $3\pi/2$.
For a fourth-rank tensor, the average over the coset operations 
is rather drastic.
The only non-zero elements of the matrix $M=(1/4)\sum_i D^4(g_i^{-1})$ 
are $M(-4,-4)=M(0,0)=M(4,4)=1$.
Therefore, the crystal-symmetrized fourth-rank tensor 
is $\langle T^4\rangle_X=\langle T^4(4)\rangle M$:
\begin{eqnarray*}
\langle T^4_0\rangle_X &=&
\sqrt{\frac{14}{5}} \langle T^4_4\rangle_X =
\sqrt{\frac{14}{5}}\langle T^4_{-4}\rangle_X=
\langle T^4_0(4)\rangle .
\end{eqnarray*}
The relation between $\langle T^4_0(4)\rangle$ and the
site-symmetrized tensor in the trigonal axes is given
by equation (\ref{T4043}).

For a second-rank tensor, the matrix $M=(1/4)\sum_i D^2(g_i^{-1})$ 
has a single non-zero element: $M(0,0)=1$.
Therefore $\langle T^2_0\rangle_X=0$, as expected \cite{BrouderAXAS}.

\subsection{The case of garnet}
We illustrate the brute force method with the case of garnet.
To calculate the spherical tensor of garnet, we use the
brute force method and calculate
$M=(1/48)\sum_R D^\ell(R)$, where the sum
runs over all the symmetry operations of the cube.

For $\ell=4$, the only nonzero matrix elements are
\begin{eqnarray*}
M(-4,-4) &=& M(-4,4)=M(4,-4)=M(4,4)= \frac{5}{24},\\
M(-4,0) &=& M(0,-4)=M(0,4)=M(4,0)= \frac{\sqrt{70}}{24},\\
M(0,0) &=& \frac{7}{12}.
\end{eqnarray*}
Therefore, the only non-zero components of a fourth-rank
tensor are
\begin{eqnarray*}
\langle T^4_0\rangle_X &=&
\sqrt{\frac{14}{5}} \langle T^4_4\rangle_X =
\sqrt{\frac{14}{5}}\langle T^4_{-4}\rangle_X=
\langle T^4_{0}(4) \rangle = 
-\frac{7 s + 2\sqrt{70} t_r}{18},
\end{eqnarray*}
where we recall that
$s=\langle T^4_0(3) \rangle$ and
$t_r=(1/2)\big(\ee^{-3i\alpha} \langle T^4_3(3)\rangle
 -\ee^{3i\alpha} \langle T^4_{-3}(3)\rangle\big)$.
Note that $s$ and $t_r$ are real because of time-reversal symmetry.

\section{Spherical average}
\label{sphavsect}
In Cartesian coordinates, the electric quadrupole matrix elements
can be written, for linearly polarized x-rays
\begin{eqnarray*}
|T_{fi}|^2 &=& \sum_{ijlm} 
  \bfepsilon_i k_j \bfepsilon_l k_m \sigma_{ijlm},
\end{eqnarray*}
with
$\sigma_{ijlm}= \langle i | x_i x_j |f \rangle \langle f |x_l x_m|i\rangle$.
For a powder sample we have the spherical average
\begin{eqnarray*}
\langle |T_{fi}|^2 \rangle &=&
\frac{1}{30}\big(2\sigma_{xxxx}+2\sigma_{yyyy}+2\sigma_{zzzz}
    +6\sigma_{xyxy} +6\sigma_{xzxz} +6\sigma_{yzyz}
\\&&
    -\sigma_{xxyy}-\sigma_{xxzz}-\sigma_{yyxx}-\sigma_{zzxx}
    -\sigma_{yyzz}-\sigma_{zzyy}
    \big).
\end{eqnarray*}
If the system is non magnetic, then
$\sigma_{iijj}=\sigma_{jjii}$ and the average further simplifies
\begin{eqnarray*}
\langle |T_{fi}|^2 \rangle &=&
\frac{1}{15}\big(\sigma_{xxxx}+\sigma_{yyyy}+\sigma_{zzzz}
    +3\sigma_{xyxy} +3\sigma_{xzxz} +3\sigma_{yzyz}
\\&&
    -\sigma_{xxyy}-\sigma_{xxzz} -\sigma_{yyzz}
    \big).
\end{eqnarray*}

\section{Formulas}

\subsection{Rotation matrix}

The rotation of $\psi$ around the direction $\bfn$
(a unit vector) is represented by the rotation matrix
(see Ref.\cite{BL}, p.~10)
$R=\id + \sin\psi N + (1-\cos\psi) N^2$,
where $N$ is the skew-symmetric matrix with
matrix elements $N_{ij}=-\sum_k \epsilon_{ijk} n_k$,
so that $(N^2)_{ij}=n_i n_j-\delta_{ij}$.

Conversely, the rotation angle $\psi$ and the rotation
axis $\bfn$ are determined from the rotation matrix $R$
by the relations
$\cos\psi=(\tr R-1)/2$,
$n_1\sin\psi=(R_{32}-R_{23})/2$,
$n_2\sin\psi=(R_{13}-R_{31})/2$,
and $n_3\sin\psi=(R_{21}-R_{12})/2$.
This is a corrected version of the relation given
in Ref.~\cite{BL} p.~20.

\subsubsection{Euler angles}
Ref.~\cite{BL}, p.~24
\begin{eqnarray*}
R = \left( \begin{array}{ccc}
               \calpha \cbeta \cgamma -\salpha\sgamma &
              -\calpha \cbeta \sgamma -\salpha\cgamma &
               \calpha\sbeta \\
               \salpha \cbeta \cgamma +\calpha\sgamma &
              -\salpha \cbeta \sgamma +\calpha\cgamma &
               \salpha\sbeta \\
              -\sbeta \cgamma &
               \sbeta \sgamma &
               \cbeta
   \end{array}\right),
\end{eqnarray*}
where $0\le\alpha < 2\pi$, $0\le\beta \le \pi$ and $0\le\gamma < 2\pi$,
$\calpha=\cos\alpha$, $\salpha=\sin\alpha$, etc.
There is a one-to-one correspondence between rotations and
parameters in this range, except for the case $\beta=0$ and
$\beta=\pi$, which describe the rotation through the angle 
$\alpha+\gamma$ and $\alpha-\gamma$, respectively, about the
axis $(0,0,1)$.

\subsubsection{Euler-Rodrigues parameters}
Although the Euler angles are more common, the
Euler-Rodrigues have the advantage of being
expressed in terms of the rotation matrix without
using circular functions. Thus, it can be convenient
to derive analytical expressions.
From the axis $\bfn$ and the angle $\psi$, we define 
the Euler-Rodrigues parameters
(Ref.\cite{BL} p.~54).
$\alpha_0=\cos(\psi/2)$, $\alpha_i=\sin(\psi/2)n_i$.
These parameters are interesting for analytic calculations
because they can be derived from the rotation matrix without
using trigonometric functions (\cite{BL} p.~19):
If $\tr R \not=-1$, then
$\alpha_0=\sqrt{\tr R +1}/2$,
$\alpha_1=(R_{32}-R_{23})/(4\alpha_0)$,
$\alpha_2=(R_{13}-R_{31})/(4\alpha_0)$,
and $\alpha_3=(R_{21}-R_{12})/(4\alpha_0)$.
If $\tr R=-1$, then
$\alpha_0=0$ and 
$\alpha_i=(\sign\alpha_i) \sqrt{(1+R_{ii})/2}$
for $i=1,2,3$ with 
$\sign\alpha_1=1$,
$\sign\alpha_2=\sign R_{12}$,
$\sign\alpha_3=\sign R_{13}$.

\begin{eqnarray*}
R = \left( \begin{array}{ccc}
               \alpha_0^2+\alpha_1^2-\alpha_2^2-\alpha_3^2 &
               2\alpha_1\alpha_2-2\alpha_0\alpha_3 &
               2\alpha_1\alpha_3+2\alpha_0\alpha_2 \\
               2\alpha_1\alpha_2+2\alpha_0\alpha_3 &
               \alpha_0^2+\alpha_2^2-\alpha_3^2-\alpha_1^2 &
               2\alpha_2\alpha_3-2\alpha_0\alpha_1 \\
               2\alpha_1\alpha_3-2\alpha_0\alpha_2 &
               2\alpha_2\alpha_3+2\alpha_0\alpha_1 &
               \alpha_0^2+\alpha_3^2-\alpha_1^2-\alpha_2^2
   \end{array}\right),
\end{eqnarray*}
where $\alpha_0=\cos(\psi/2)$, $\alpha_i=\sin(\psi/2)n_i$.

\subsection{Solid harmonics}
\label{solharmsect}
For a vector $\bfr=(x,y,z)$, the solid harmonics
$Y_\ell^m(\bfr)$ are defined by (see Ref.~\cite{BL} p.~71)
\begin{eqnarray*}
Y_\ell^m(\bfr) &=&
\sqrt{\frac{(2\ell+1)(\ell+m)!(\ell-m)!}{4\pi}}
\sum_{k} \frac{(-x-iy)^{k+m}(x+iy)^k
z^{\ell-2k-m}}{2^{2k+m}(k+m)!k!(l-m-2k)!},
\end{eqnarray*}
where $k$ runs from $\max(0,-m)$ to the integer part of
$(\ell-m)/2$.
The most important example of solid harmonics is
\begin{eqnarray*}
Y_1(\bfr) &=& \sqrt{\frac{3}{4\pi}}
   \left( \begin{array}{c} 
           \frac{x-iy}{\sqrt{2}} \\
           z \\
          -\frac{x+iy}{\sqrt{2}}
   \end{array}\right),
\end{eqnarray*}
where the upper component is $Y_1^{-1}(\bfr)$.

\subsection{Wigner matrices}
There are several representations of the 
Wigner rotation matrices. We present here
the expressions in terms of Euler angles and
of Euler-Rodrigues parameters. Other formulas
have been derived, for example the recent 
invariant spinor representation \cite{Manakov}.

\subsubsection{Euler angles}
\label{EulerWignersect}
For a rotation $R$ expressed in terms of Euler angles
$\alpha,\beta,\gamma$, the Wigner matrix is
(\cite{BL} p.~46)
\begin{eqnarray*}
D^\ell_{m'm}(R) &=& \ee^{-im'\alpha}d^\ell_{m'm}(\beta)\ee^{-im\gamma}.
\end{eqnarray*}
Various expressions exist for the
reduced Wigner matrix $d^\ell_{m'm}(\beta)$.
The following formula (valid for half-integer $\ell$)
is particularly convenient for computers,
because it avoids the presence of singular terms
(\cite{BL} p.~50):
\begin{eqnarray*}
d^\ell_{m'm}(\beta) &=&
(-1)^\lambda \sqrt{\frac{k!(2\ell-k)!}{(k+\mu)!(k+\nu)!}}
  \big(\sin\frac{\beta}{2}\big)^{\mu}
  \big(\cos\frac{\beta}{2}\big)^{\nu}
  P^{(\mu,\nu)}_k(\cos\beta),
\end{eqnarray*}
where $k=\min(\ell+m,\ell-m,\ell+m',\ell-m')$ and
the non-negative integers
$\mu$, $\nu$ and $\lambda$ are determined by the value of $k$
\begin{itemize}
\item if $k=\ell+m$, then $\mu=m'-m$, $\nu=-m'-m$ and $\lambda=m'-m'$,
\item if $k=\ell-m$, then $\mu=m-m'$, $\nu=m'+m$ and $\lambda=0$,
\item if $k=\ell+m'$, then $\mu=m-m'$, $\nu=-m'-m$ and $\lambda=0$,
\item if $k=\ell-m'$, then $\mu=m'-m$, $\nu=-m'+m$ and $\lambda=m'-m'$.
\end{itemize}
In this expression, the only possible numerical difficulty 
occurs for $0^0$ that should be 1.
The Jacobi polynomials $P^{(\mu,\nu)}_k(x)$ are given by the
formula
\begin{eqnarray*}
P^{(\mu,\nu)}_k(x) &=& \sum_{i=0}^k 
   {k+\mu\choose i} 
   {k+\nu\choose k-i} 
   \left(\frac{x-1}{2}\right)^{k-i} \left(\frac{x+1}{2}\right)^i.
\end{eqnarray*}

For example, the Wigner matrix for first-rank
tensors is
\begin{eqnarray*}
D^1 = \left( \begin{array}{ccc}
               \frac{\cos\beta+1}{2}\ee^{i(\alpha+\gamma)} &
               \frac{\sin\beta}{\sqrt{2}}\ee^{i\alpha} &
               \frac{\cos\beta-1}{2}\ee^{i(\alpha-\gamma)} \\
              -\frac{\sin\beta}{\sqrt{2}}\ee^{i\gamma} &
               \cos\beta &
               \frac{\sin\beta}{\sqrt{2}}\ee^{-i\gamma} \\
               \frac{\cos\beta-1}{2}\ee^{i(\gamma-\alpha)} &
              -\frac{\sin\beta}{\sqrt{2}}\ee^{-i\alpha} &
               \frac{\cos\beta+1}{2}\ee^{-i(\alpha+\gamma)}
   \end{array}\right),
\end{eqnarray*}
where the upper left matrix element is $D^1_{-1-1}$. Two useful special
cases are $d^\ell_{mm'}(0)=\delta_{mm'}$
and $d^\ell_{mm'}(\pi)=(-1)^{\ell+m}\delta_{m,-m'}$.

\subsubsection{Euler-Rodrigues parameters}

The Wigner rotation matrix is
\begin{eqnarray*}
D^\ell_{m'm}(R) &=&
\sqrt{(\ell+m')!(\ell-m')!(\ell+m)!(\ell-m)!}
\\&&\hspace*{-16mm}
\sum_{k}
\frac{(\alpha_0-i\alpha_3)^{\ell+m-k}(-i\alpha_1-\alpha_2)^{m'-m+k}
  (-i\alpha_1+\alpha_2)^k (\alpha_0+i\alpha_3)^{\ell-m'-k}}
  {(\ell+m-k)!(m'-m+k)!k!(\ell-m'-k)!},
\end{eqnarray*}
where $k$ runs from $\max(0,m-m')$ to $\min(\ell+m,\ell-m')$.

\subsection{Butler's orientation}
\label{Butlersect}
The powerful multiplet program developed by Theo Thole and
Barry Searle is based on Butler's conventions.
For the calculation of trigonal sites in cubic crystals,
it is necessary to know precisely the relation between the cubic 
and trigonal reference frames, which is not clearly stated in 
Butler's book. To determine it,
we combine Butler's tables pp.~522, 527 and 549 of \cite{Butler}.
This shows that the transition between spherical harmonics 
$|1m\rangle_3$ in the trigonal axes (i.e. in the $O$-$D_3$-$C_3$ basis)
and spherical harmonics $|1m\rangle_4$ in the cubic axes
(i.e.  in the $O$-$D_4$-$C_4$ basis) is 
\begin{eqnarray*}
|1-1\rangle_3 &=& 
   |1-1\rangle_4 \frac{(1-i)(\sqrt{3}+1)}{\sqrt{24}} + 
    |10\rangle_4\frac{1}{\sqrt{3}}
   + |11\rangle_4 \frac{(1+i)(\sqrt{3}-1)}{\sqrt{24}},\\
|10\rangle_3 &=& 
   |1-1\rangle_4 \frac{-1+i}{\sqrt{6}} + |10\rangle_4 \frac{1}{\sqrt{3}} 
   + |11\rangle_4 \frac{1+i}{\sqrt{6}},\\
|11\rangle_3 &=& 
   |1-1\rangle_4 \frac{(1-i)(\sqrt{3}-1)}{\sqrt{24}} + 
     |10\rangle_4\frac{-1}{\sqrt{3}} 
   + |11\rangle_4 \frac{(1+i)(\sqrt{3}+1)}{\sqrt{24}}.
\end{eqnarray*}
This can be rewritten
\begin{eqnarray*}
|1m\rangle_3 &=& \sum_{m'=-1}^1 |1m'\rangle_4
      D^1_{m'm}(R),
\end{eqnarray*}
for the rotation $R$ corresponding to the Euler angles
$\alpha=3\pi/4$, $\beta=\arccos(1/\sqrt{3})$ and $\gamma=\pi$.
This corresponds to the $C_{3z}$ axis of $D_3$ along the (-1,1,1) 
direction of the cube and the $C_{2y}$ axis of $D_3$
along the (1,1,0) direction of the cube
(see fig.~11.6 of Ref.~\cite{Butler}, p.~204).
The inverse rotation has Euler angles $(0,\beta_0,\pi/4)$.
To be still more detailed, the rotation
\begin{eqnarray}
R=R(3\pi/4,\beta_0,\pi) &=& \left( \begin{array}{ccc}
               1/\sqrt{6} &
               1/\sqrt{2} &
              -1/\sqrt{3} \\
              -1/\sqrt{6} &
               1/\sqrt{2} &
               1/\sqrt{3} \\
               \sqrt{2/3} &
               0 &
               1/\sqrt{3}
   \end{array}\right),
\label{Butler43}
\end{eqnarray}
transforms any symmetry operation $R'$ in the $D_{3d}$ axes into 
the symmetry operation $R R' R^{-1}$ in the cubic axes.

\section{Coupling identities}
\label{couplingidsect}
We gather some useful coupling formulas.
If $\bfa$, $\bfb$, $\bfc$ and $\bfd$ are vectors, we denote
by $\bfa^1$, $\bfb^1$, $\bfc^1$ and $\bfd^1$ the corresponding first-rank
spherical tensors. Then, according to
reference \cite{VMK} p.~66 and 67,
\begin{eqnarray}
\{\bfa^1\potimes\bfb^1\}^1&=& \frac{i}{\sqrt{2}} (\bfa\times\bfb)^1.
\label{vecprod}
\end{eqnarray}
\begin{eqnarray}
\{\{\bfa^1\potimes\bfb^1\}^0\potimes \{\bfc^1\potimes\bfd^1\}^0\}^0
&=& \frac{1}{3} (\bfa\cdot\bfb)(\bfc\cdot\bfd).
\label{prod0}
\end{eqnarray}
\begin{eqnarray}
\{\{\bfa^1\potimes\bfb^1\}^2\potimes \{\bfc^1\potimes\bfd^1\}^2\}^0
&=& \frac{1}{\sqrt{5}} 
  \big(\frac{1}{2} (\bfa\cdot\bfc)(\bfb\cdot\bfd) 
  +    \frac{1}{2} (\bfa\cdot\bfd)(\bfb\cdot\bfc) 
\nonumber\\&&
  -    \frac{1}{2} (\bfa\cdot\bfb)(\bfc\cdot\bfd) \big).
\label{prod4}
\end{eqnarray}

To prove equation (\ref{recouple0}), we start from the identity
\begin{eqnarray*}
\{P^a\potimes Q^b\}^c \cdot \{R^d\potimes S^e\}^c
&=&
(-1)^{2a+b-d} \sum_g (2c+1) \sixj{a}{b}{c}{e}{d}{g}
\\&&\times
\{P^a\potimes R^d\}^g \cdot \{Q^b\potimes S^e\}^g,
\end{eqnarray*}
where $g$ runs between $\max(|a-d|,|b-e|)$ and $\min(a+d,b+e)$
(see equation (13) p.~70 of \cite{VMK}).
Equation (\ref{recouple0}) corresponds to the case $c=0$
because of the special value of the
$6j$-symbol (equation (1) p.~299 of \cite{VMK})
\begin{eqnarray*}
\sixj{a}{b}{0}{e}{d}{g} &=& (-1)^{a+e+f} 
    \frac{\delta_{ab}\delta_{de}}{\sqrt{(2a+1)(2d+1)}}.
\end{eqnarray*}

The interplay between Wigner matrices and Clebsch-Gordan coefficients
is described by the following identity
(equation (5) p.~85 of \cite{VMK})
\begin{eqnarray}
\sum_{\gamma'} (a\alpha b\beta|c\gamma') D^c_{\gamma'\gamma}(R) 
&=& 
\sum_{\alpha'\beta'} (a\alpha' b\beta'|c\gamma) 
D^a_{\alpha\alpha'}(R) D^c_{\beta\beta'}(R).
\label{WignerCG}
\end{eqnarray}

\ack
We thank Etienne Balan and Philippe Sainctavit for very constructive comments.

\section*{References}
%\bibliographystyle{jpa}
%\bibliography{qed}% Produces the bibliography via BibTeX.

\begin{thebibliography}{10}

\bibitem{Juhin}
Juhin A, Brouder C, Arrio M~A, Cabaret D, Sainctavit P, Balan E, Bordage A,
  Calas G, Eeckhout S~G and Glatzel P 2008 X-ray natural dichroism in cubic
  compounds: the case of substitutional {Cr}$^{3+}$ in {MgAl$_2$O$_4$} To be
  published

\bibitem{BL}
Biedenharn L and Louck J 1981 \emph{Angular Momentum in Quantum Physics} vol.~8
  of \emph{Encyclopedia of Mathematics and its Applications} (Reading:
  Addison-Wesley)

\bibitem{VMK}
Varshalovich D~A, Moskalev A~N and Khersonskii V~K 1988 \emph{Quantum Theory of
  Angular Momentum} (Singapore: World Scientific)

\bibitem{GoulonM}
Goulon J 1989 Syst\`emes mol\'eculaires : {D}ichro\"{\i}sme circulaire naturel
  et magn\'etique dans les spectroscopies optiques ou d'excitation des couches
  internes in E~Beaurepaire, B~Carri\`ere and J~P Kappler, eds.,
  \emph{Rayonnement synchrotron polaris\'e, \'electrons polaris\'es et
  magn\'etisme} (Strasbourg: IPCMS) pp. 333--86

\bibitem{BrouderNCD}
Natoli C, Brouder C, Sainctavit P, Goulon J, Goulon-Ginet C and Rogalev A 1998
  Calculation of x-ray natural circular dichroism \emph{Europ. Phys. J. B}
  \textbf{4} 1--11

\bibitem{Carra00}
Carra P and Benoist R~R 2000 X-ray natural circular dichroism \emph{Phys. Rev.
  B} \textbf{62} R7703--6

\bibitem{LudwigFalter}
Ludwig W and Falter C 1996 \emph{Symmetries in Physics: Group Theory applied to
  Physical Problems} 2nd ed. (Berlin: Springer)

\bibitem{Carra03}
Carra P, Jerez A and Marri I 2003 X-ray dichroism in noncentrosymmetric
  crystals \emph{Phys. Rev. B} \textbf{67} 045111

\bibitem{MarriCarra}
Marri I and Carra P 2004 Scattering operators for {E1-E2} x-ray resonant
  diffraction \emph{Phys. Rev. B} \textbf{69} 113101

\bibitem{BrouderAXAS}
Brouder C 1990 Angular dependence of x-ray absorption spectra \emph{J. Phys.:
  Condens. Matter} \textbf{2} 701--38

\bibitem{Cowan}
Cowan R~D 1981 \emph{The Theory of Atomic Structure and Spectra} (Berkeley:
  University of Calivornia Press)

\bibitem{Butler}
Butler P~H 1981 \emph{Point Symmetry Group Applications} (New York: Plenum
  Press)

\bibitem{Thole85}
Thole B~T, van~der Laan G, Fuggle J~C, Sawatzky G~A, Karnatak R~C and Esteva
  J~M 1085 $3d$ x-ray-absorption lines and the $3d^9 4f^{n+1}$ multiplets of
  lanthanides \emph{Phys. Rev. B} \textbf{32} 5107--18

\bibitem{Kotani}
Kotani A, Ogasawara H, Okada K, Thole B~T and Sawatzky G~A 1989 Theory of
  multiplet structure in 4d core photoabsorption spectra of {CeO$_2$}
  \emph{Phys. Rev. B} \textbf{40} 65--73

\bibitem{Kuiper}
Kuiper P, Searle B~G, Rudolf P, Tjeng L~H and Chen C~T 1957 X-ray magnetic
  dichroism and antiferromagnetic {F}e$_2${O}$_3$: {T}he orientation of
  magnetic moments observed by {F}e $2p$ x-ray-absorption spectroscopy
  \emph{Phys. Rev. Lett.} \textbf{70} 1549--52

\bibitem{Hahn}
Hahn T 2002 \emph{International Tables for Crystallography: Volume A} 5th ed.
  (Dordrecht: Kluwer Academic Publishers)

\bibitem{Bordage}
Bordage A and coll 2008 Substitutional chromium in garnet In preparation

\bibitem{Manakov}
Manakov N~L, Meremianin A~V and Starace A~F 2001 Invariant spinor
  representations of finite rotation matrices \emph{Phys. Rev. A} \textbf{64}
  032105

\end{thebibliography}

\end{document}